\documentclass[10pt]{article}
\usepackage[utf8]{inputenc}
\usepackage{amsmath}
\usepackage{amssymb}
\usepackage{graphicx}
\usepackage[table]{xcolor}
\usepackage{orcidlink}
\usepackage{hyperref}
\usepackage{graphicx}
\usepackage{epstopdf}
\usepackage{amsmath}
\usepackage{enumitem}
\usepackage{epsfig}
\usepackage{longtable}
\hypersetup{colorlinks=true, linkcolor=blue, citecolor=blue, urlcolor=blue}
\usepackage{caption}
\usepackage{subcaption}
\usepackage{subcaption}
\RequirePackage[numbers,sort&compress]{natbib}
\begin{document}
\baselineskip=16pt

\newcommand{\apjl}{APJL}
\newcommand{\apj}{APJ}
\newcommand{\aap}{AAP}
\newcommand{\mnras}{MNRAS}
\newcommand{\nar}{NAR}
\newcommand{\apjs}{APJS}
\newcommand{\jcap}{JCAP}
\newcommand{\prd}{PRD}
\newcommand{\apss}{APSS}
\newcommand{\prl}{PRL}
\newcommand{\nat}{NAT}
\newcommand{\araa}{ARAA}
\newcommand{\pasj}{PASJ}
\newcommand{\plb}{PLB}

\begin{center}
\LARGE{Analytic and Numerical Constraints on QPOs in EHT and XRB Sources Using Quantum-Corrected Black Holes}
\end{center}

\vspace{0.3cm}
\begin{center}
{\bf Ahmad Al-Badawi\orcidlink{0000-0002-3127-3453}}\footnote{\bf ahmadbadawi@ahu.edu.jo}\\
{\it Department of Physics, Al-Hussein Bin Talal University 71111, Ma’an, Jordan}\\
{\bf Faizuddin Ahmed\orcidlink{0000-0003-2196-9622}}\footnote{\bf faizuddinahmed15@gmail.com}\\
{\it Department of Physics, Royal Global University, Guwahati, 781035, Assam, India}\\
{\bf Orhan Dönmez\orcidlink{0000-0001-9017-2452}}\footnote{\bf orhan.donmez@aum.edu.kw (Corresp. author)}\\
{\it College of Engineering and Technology, American University of the Middle East, Egaila 54200, Kuwait}\\
{\bf Fatih Doğan\orcidlink{0000-0001-7992-1723}}\footnote{\bf fatih.dogan@aum.edu.kw}\\
{\it College of Engineering and Technology, American University of the Middle East, Egaila 54200, Kuwait}\\
{\bf Behnam Pourhassan\orcidlink{xxx}}\footnote{\bf b.pourhassan@du.ac.ir}\\
{\it School of Physics, Damghan University, Damghan 3671645667, Iran}\\
{\it Center for Theoretical Physics, Khazar University, 41 Mehseti Street, Baku, AZ1096, Azerbaijan}\\
{\bf \.{I}zzet Sakall{\i}\orcidlink{0000-0001-7827-9476}}\footnote{\bf izzet.sakalli@emu.edu.tr}\\
{\it Physics Department, Eastern Mediterranean University, Famagusta 99628, North Cyprus via Mersin 10, Turkey}\\
{\bf Yassine Sekhmani\orcidlink{xxxx}}\footnote{\bf sekhmaniyassine@gmail.com}\\
{\it Center for Theoretical Physics, Khazar University, 41 Mehseti Street, Baku, AZ1096, Azerbaijan}\\
{\it Centre for Research Impact \& Outcome, Chitkara University Institute of Engineering and Technology,
Chitkara University, Rajpura, 140401, Punjab, India}
\end{center}

\vspace{0.3cm}

\begin{abstract}
This investigation examines quasi-periodic oscillations (QPOs) in two quantum-corrected black hole (BH) spacetimes that preserve general covariance while incorporating quantum gravitational effects through a dimensionless parameter $\zeta$. We combine analytical derivations of epicyclic frequencies with comprehensive numerical simulations of Bondi-Hoyle-Lyttleton (BHL) accretion to explore how quantum corrections manifest in observable astrophysical phenomena. Using a fiducial BH mass of $M=10M_\odot$ representative of stellar-mass X-ray binaries, we demonstrate that the two models exhibit fundamentally different behaviors: Model-I modifies both temporal and radial metric components, leading to innermost stable circular orbit migration proportional to $\zeta^4$ and dramatic stagnation point evolution from $27M$ to $5M$ as quantum corrections strengthen. Model-II preserves the classical temporal component while altering only spatial geometry, maintaining constant stagnation points and stable cavity structures throughout the parameter range. Our numerical simulations reveal distinct QPO generation mechanisms, with Model-I showing systematic frequency evolution and cavity shrinkage that suppresses oscillations for $\zeta \geq 3M$, while Model-II maintains stable low-frequency modes up to $\zeta \geq 5M$. Power spectral density analyzes demonstrate characteristic frequency ratios ($3:2$, $2:1$, $5:3$) consistent with observations from X-ray binaries, providing specific targets for discriminating between quantum correction scenarios. The hydrodynamically derived constraints ($\zeta \lesssim 4M$) show remarkable agreement with independent Event Horizon Telescope limits for M87$^*$ and Sgr A$^*$, validating our theoretical framework through multiple observational channels. These results establish QPO frequency analysis as a probe for detecting quantum gravitational effects in astrophysical BHs and demonstrate the complementary nature of timing and imaging observations in constraining fundamental physics.
\end{abstract}



\section{Introduction}
\label{isec1}

QPOs represent one of the most compelling observational phenomena in high-energy astrophysics, providing direct probes into the strong gravitational field regime surrounding compact objects \cite{isz01,isz02,isz03}. These characteristic frequency signatures, observed in the X-ray light curves of accreting BH and neutron stars, encode fundamental information about spacetime geometry, orbital dynamics, and the underlying physics governing matter behavior in extreme environments \cite{isz04,isz05}. The analytical and numerical analysis of QPO frequencies has emerged as a powerful diagnostic tool for testing theories of gravity, constraining BH parameters, and potentially detecting signatures of physics beyond General Relativity (GR) \cite{isz06,isz07,isz08,Orh6,isz54}.

The theoretical framework for understanding QPOs is intimately connected to the epicyclic motion of test particles in curved spacetime, where small perturbations around stable circular orbits give rise to characteristic radial, vertical, and azimuthal oscillation frequencies \cite{isz09,isz10}. In the Schwarzschild and Kerr metrics of classical GR, these frequencies exhibit well-defined relationships that have been extensively studied and compared with observations from X-ray binaries (XRBs) \cite{isz11,isz12}. The radial epicyclic frequency $\nu_r$, vertical epicyclic frequency $\nu_\theta$, and orbital frequency $\nu_\phi$ form the basis for various QPO models, including the relativistic precession model and diskoseismic oscillations \cite{isz13,isz14}. Observational evidence for commensurate frequency ratios, particularly the ubiquitous $3:2$ ratio observed in high-frequency QPOs (HFQPOs) and the diverse patterns in low-frequency QPOs (LFQPOs), has motivated extensive theoretical investigations into the underlying mechanisms \cite{isz15,isz16,isz17}.

The advent of Event Horizon Telescope (EHT) observations of supermassive BH, particularly M87$^*$ and Sgr A$^*$, has revolutionized our understanding of  BH physics and opened new avenues for testing alternative theories of gravity \cite{isz18,isz19}. These groundbreaking observations provide unprecedented constraints on BH shadows, photon rings, and near-horizon dynamics, establishing stringent limits on deviations from classical GR predictions \cite{isz20,isz51}. Simultaneously, the detection of gravitational waves by the Laser Interferometer Gravitational-Wave Observatory (LIGO) and Virgo collaborations has further emphasized the importance of understanding strong-field gravity and motivating searches for quantum gravitational effects \cite{isz22,isz23}.

Within this context, quantum-corrected black hole (QCBH) solutions have gained significant attention as natural extensions of classical GR that incorporate quantum gravitational effects \cite{Sucu:2025rux,Sucu:2025fwa,Ahmed:2025qor,Ahmed:2025iqz,Ahmed:2025evv,Pourhassan:2025bth}
 while preserving general covariance \cite{isz24,isz25}. These models arise from effective quantum gravity theories and introduce modifications to the spacetime metric through dimensionless quantum correction parameters, typically denoted as $\zeta$ \cite{isz26,isz27}. Unlike ad hoc modifications to Einstein's equations, QCBHs maintain the fundamental principles of general covariance and reduce to classical solutions in appropriate limits, making them theoretically well-motivated candidates for studying quantum gravity phenomenology \cite{isz28,isz29}.

The investigation of QPO frequencies in QCBH spacetimes represents a natural intersection of theoretical quantum gravity and observational astrophysics. Analytical studies of epicyclic motion in modified spacetimes have revealed that quantum corrections can significantly alter the characteristic frequencies, potentially leading to observable deviations from classical GR predictions \cite{isz30,isz31}. However, the complexity of these modifications necessitates comprehensive numerical investigations to fully understand their implications for realistic astrophysical scenarios. The BHL accretion mechanism provides a particularly relevant framework for studying matter dynamics around compact objects, as it naturally generates shock cone structures that can trap and amplify oscillatory modes \cite{isz32,isz33,Orh3}.

Our investigation focuses on two distinct QCBH models that preserve general covariance while incorporating quantum corrections in fundamentally different ways. Model-I modifies both the temporal and radial metric components through quantum corrections, leading to coupled modifications in energy and angular momentum conservation. In contrast, Model-II preserves the classical temporal component while introducing quantum corrections only in the radial metric function, maintaining the standard redshift relationships while altering spatial geometry \cite{isz34,isz35}. This distinction proves crucial for understanding how different types of quantum corrections manifest in observable phenomena, particularly in the context of QPO generation and evolution. Throughout this work, we adopt a fiducial BH mass of $M=10M_\odot$ for numerical calculations, which is representative of stellar-mass BH in X-ray binaries and allows direct comparison with observational data from well-studied sources.

The motivation for this work stems from several converging factors in contemporary astrophysics and fundamental physics. First, the increasing precision of X-ray timing observations, exemplified by missions such as the Neutron Star Interior Composition Explorer (NICER) and the upcoming enhanced X-ray Timing and Polarimetry (eXTP) mission, demands theoretical frameworks capable of predicting subtle deviations from classical GR \cite{isz36,isz37}. Second, the success of EHT observations in constraining near-horizon physics provides complementary constraints that can be combined with QPO studies to probe quantum gravitational effects across multiple observational channels \cite{isz38,isz39}. Third, the ongoing search for observational signatures of quantum gravity requires systematic investigations of how quantum corrections manifest in different astrophysical contexts, from individual particle orbits to collective hydrodynamic phenomena.

The primary aims of this investigation are fourfold. First, we derive comprehensive analytical expressions for epicyclic frequencies in both QCBH models, establishing the theoretical foundation for understanding quantum corrections to orbital dynamics. These analytical results provide crucial insights into the parameter dependence of frequency modifications and identify the regimes where quantum effects become observationally significant. Second, we perform high-resolution numerical simulations of BHL accretion around QCBHs, employing relativistic hydrodynamic codes to capture the full nonlinear dynamics of matter flow, shock formation, and instability development. These simulations reveal the complex interplay between modified spacetime geometry and collective matter behavior, demonstrating how quantum corrections influence macroscopic astrophysical phenomena.

Third, we conduct detailed power spectral density (PSD) analyses of the numerical simulation outputs to extract characteristic QPO frequencies and compare them with both analytical predictions and observational data from known XRB sources. This comparison establishes the observational viability of QCBH models and identifies specific frequency signatures that could distinguish quantum-corrected solutions from classical GR. The use of $M=10M_\odot$ as our reference mass scale enables direct frequency comparisons with observations from stellar-mass BH systems, while our scaling relationships allow extrapolation to supermassive BH. Fourth, we investigate the consistency between our hydrodynamical constraints on the quantum correction parameter $\zeta$ and independent limits derived from EHT observations, demonstrating the complementary nature of different observational probes in constraining quantum gravitational effects.

The paper is organized as follows: Section~\ref{isec2} presents the theoretical framework for QCBH spacetimes and analyzes their horizon structures, establishing the geometric foundation for subsequent investigations. Section~\ref{isec3} derives analytical expressions for epicyclic frequencies and periastron precession in both QCBH models, providing comprehensive treatment of orbital dynamics modifications. Section~\ref{isec4} describes our numerical BHL accretion simulations, analyzing shock cone morphology, stagnation point behavior, and QPO generation mechanisms in quantum-corrected spacetimes. Section~\ref{isec5} presents systematic comparisons between numerical results, analytical predictions, and observational constraints from both XRB timing studies and EHT observations. Finally, Section~\ref{isec6} summarizes our main conclusions and discusses implications for future theoretical and observational investigations of quantum gravity effects in astrophysical BH. Unless otherwise stated, geometrized units with $G = c = 1$ have been used throughout this paper.

\section{QCBH Spacetimes and Horizon Structure} \label{isec2}

The theoretical framework for our investigation centers on two distinct QCBH solutions that emerge from effective quantum gravity theories \cite{isz25,isz26}. These models incorporate quantum corrections through a dimensionless parameter $\zeta$ while preserving general covariance, representing significant departures from classical GR in the strong-field regime \cite{isz27,isz24}.

The general spherically symmetric metric for both QCBH models takes the form:
\begin{equation}
    ds^2 = -f(r)\, dt^2 + \frac{1}{g(r)}\, dr^2 + r^2\,\left(d\theta^2 + \sin^2\theta\, d\phi^2\right),\label{metric}
\end{equation}
where the metric functions $f(r)$ and $g(r)$ distinguish the two models under consideration.

For Model-I, both temporal and radial metric components are modified by quantum corrections \cite{QPOmetric}:
\begin{align}
f(r) &=1 -\frac{2M}{r}+ \frac{\zeta^2}{r^2}\,\left(1- \frac{2M}{r}\right)^2, \nonumber \\
g(r) &=1 -\frac{2M}{r}+ \frac{\zeta^2}{r^2}\,\left(1- \frac{2M}{r}\right)^2, \qquad  \text{\bf Model-I}. \label{function-1}
\end{align}

For Model-II, only the radial metric component incorporates quantum corrections while the temporal component remains Schwarzschild-like \cite{QPOmetric}:
\begin{align}
f(r) &=1 -\frac{2M}{r}, \nonumber \\
g(r) &= 1 -\frac{2M}{r}+ \frac{\zeta^2}{r^2}\,\left(1 -\frac{2M}{r}\right)^2, \qquad  \text{\bf Model-II}. \label{function-2}
\end{align}

Here, $M$ represents the ADM mass of the BH, and $\zeta$ is the quantum correction parameter with dimensions of length \cite{sec2is05}. The classical limit $\zeta \rightarrow 0$ recovers the standard Schwarzschild solution for both models, ensuring consistency with GR in the absence of quantum effects \cite{sec2is06}.

A crucial observation is that the horizon structure for both models is governed by the identical function $g(r)$, meaning the event horizon location $r_h$ is determined by solving $g(r_h) = 0$ for both Model-I and Model-II. This shared horizon condition yields identical causal structures despite the different temporal metric behaviors \cite{sec2is07}.

The horizon analysis reveals rich causal structure depending on the quantum parameter $\zeta$. As demonstrated in Table~\ref{horizons_model1_2}, the spacetime undergoes a fundamental transition from the classical Schwarzschild configuration to a more complex two-horizon structure as $\zeta$ increases from zero.

\setlength{\tabcolsep}{12pt}
\begin{longtable}{|c|c|c|}
\hline
\rowcolor{orange!50}
\textbf{$\zeta$} & \textbf{Horizon(s)} & \textbf{Configuration} \\
\hline
\endfirsthead
\hline
\rowcolor{orange!50}
\textbf{$\zeta$} & \textbf{Horizon(s)} & \textbf{Configuration} \\
\hline
\endhead
0.0 & [2.0] & Extremal or Single Root BH \\
\hline
0.1 & [0.25917041, 2.0] & Non-extremal BH \\
\hline
0.5 & [0.68939835, 2.0] & Non-extremal BH \\
\hline
1.0 & [1.0, 2.0] & Non-extremal BH \\
\hline
1.5 & [1.2108937, 2.0] & Non-extremal BH \\
\hline
2.0 & [1.3646556, 2.0] & Non-extremal BH \\
\hline
2.5 & [1.4806400, 2.0] & Non-extremal BH \\
\hline
3.0 & [1.5700065, 2.0] & Non-extremal BH \\
\hline
4.0 & [1.6954152, 2.0] & Non-extremal BH \\
\hline
5.0 & [1.7759473, 2.0] & Non-extremal BH \\
\hline
\caption{\footnotesize Horizons of the quantum-corrected BH (Model-I and Model-II) with metric function $g(r) = 1 - \frac{2M}{r} + \frac{\zeta^2}{r^2} \left( 1 - \frac{2M}{r} \right)^2$, where $M = 1$ is fixed, and $\zeta \in \{0.0, 0.1, 0.5, 1.0, 1.5, 2.0, 2.5, 3.0, 4.0, 5.0\}$ varies. The spacetime exhibits a transition from an extremal or single-root BH at $\zeta = 0.0$ (Schwarzschild limit with $r_h = 2M$) to non-extremal BH with two horizons for $\zeta > 0$, where the inner horizon increases with $\zeta$ while the outer horizon remains fixed at $r_h = 2M$.}
\label{horizons_model1_2}
\end{longtable}

The data in Table~\ref{horizons_model1_2} reveals several important features of the QCBH horizon structure. For $\zeta = 0$, we recover the familiar Schwarzschild configuration with a single horizon at $r_h = 2M = 2.0$ (in units where $M = 1$). However, any non-zero quantum correction parameter $\zeta > 0$ immediately generates a two-horizon structure characteristic of non-extremal BH. Remarkably, while the outer horizon remains fixed at $r_h = 2M$ regardless of $\zeta$, the inner horizon position increases monotonically with the quantum parameter, ranging from $r_{inner} \approx 0.259M$ for $\zeta = 0.1$ to $r_{inner} \approx 1.776M$ for $\zeta = 5.0$.

This behavior suggests that quantum corrections effectively "push" the inner horizon outward while preserving the classical outer horizon location, fundamentally altering the BH's internal causal structure \cite{sec2is08}. The physical interpretation involves quantum effects modifying the spacetime geometry in a way that creates an additional trapped surface inside the classical Schwarzschild radius.

\begin{figure}[ht!]
    \centering
    \setlength{\tabcolsep}{0pt} 

    \begin{minipage}{0.32\textwidth}
        \centering
        \includegraphics[width=\textwidth]{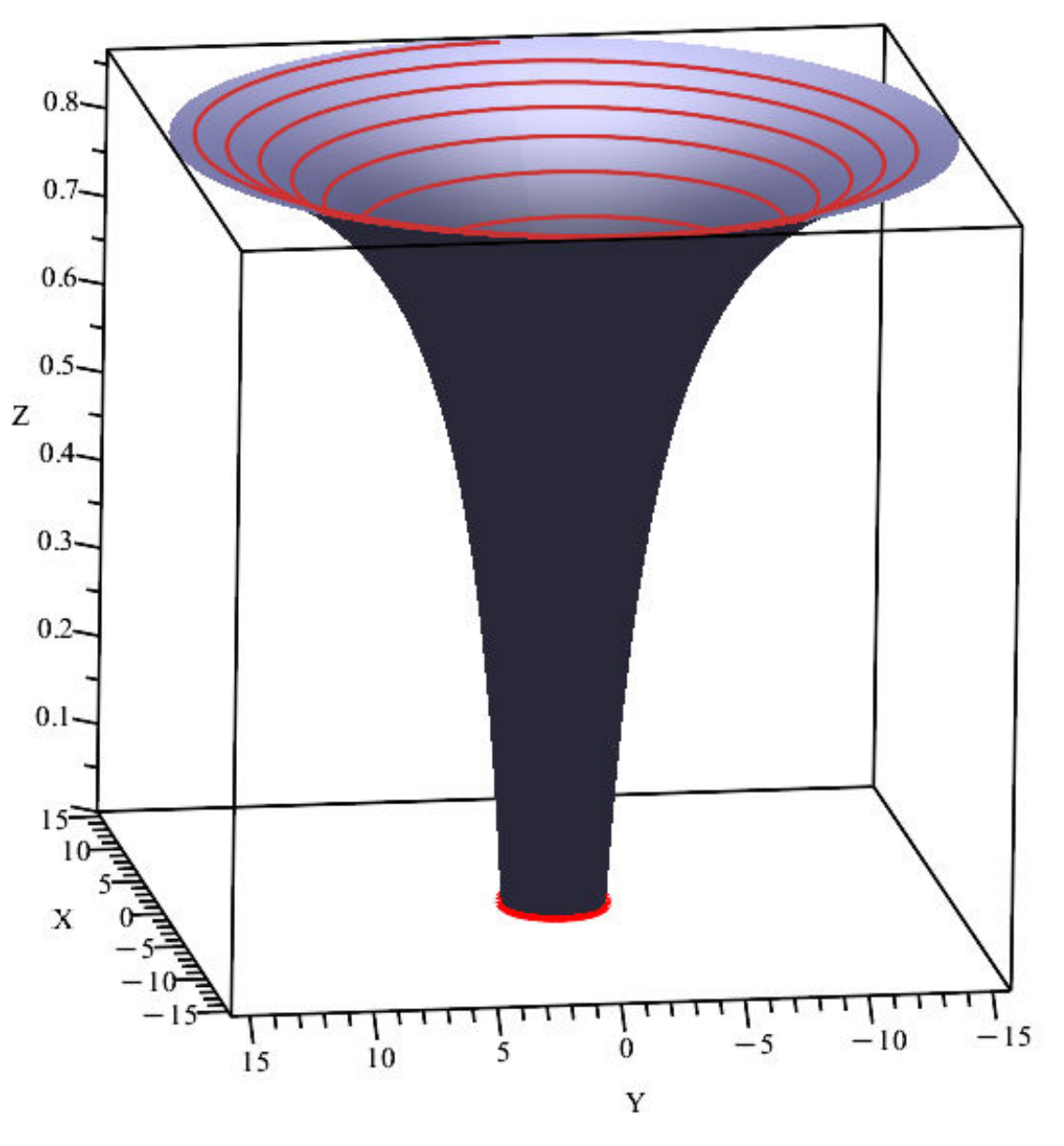}
        \subcaption{[$\zeta=0$, $M=1$, $r_h=2.0$] \newline Extremal or Single Root BH (Schwarzschild limit).}
        \label{fig:zeta0}
    \end{minipage}
    \begin{minipage}{0.32\textwidth}
        \centering
        \includegraphics[width=\textwidth]{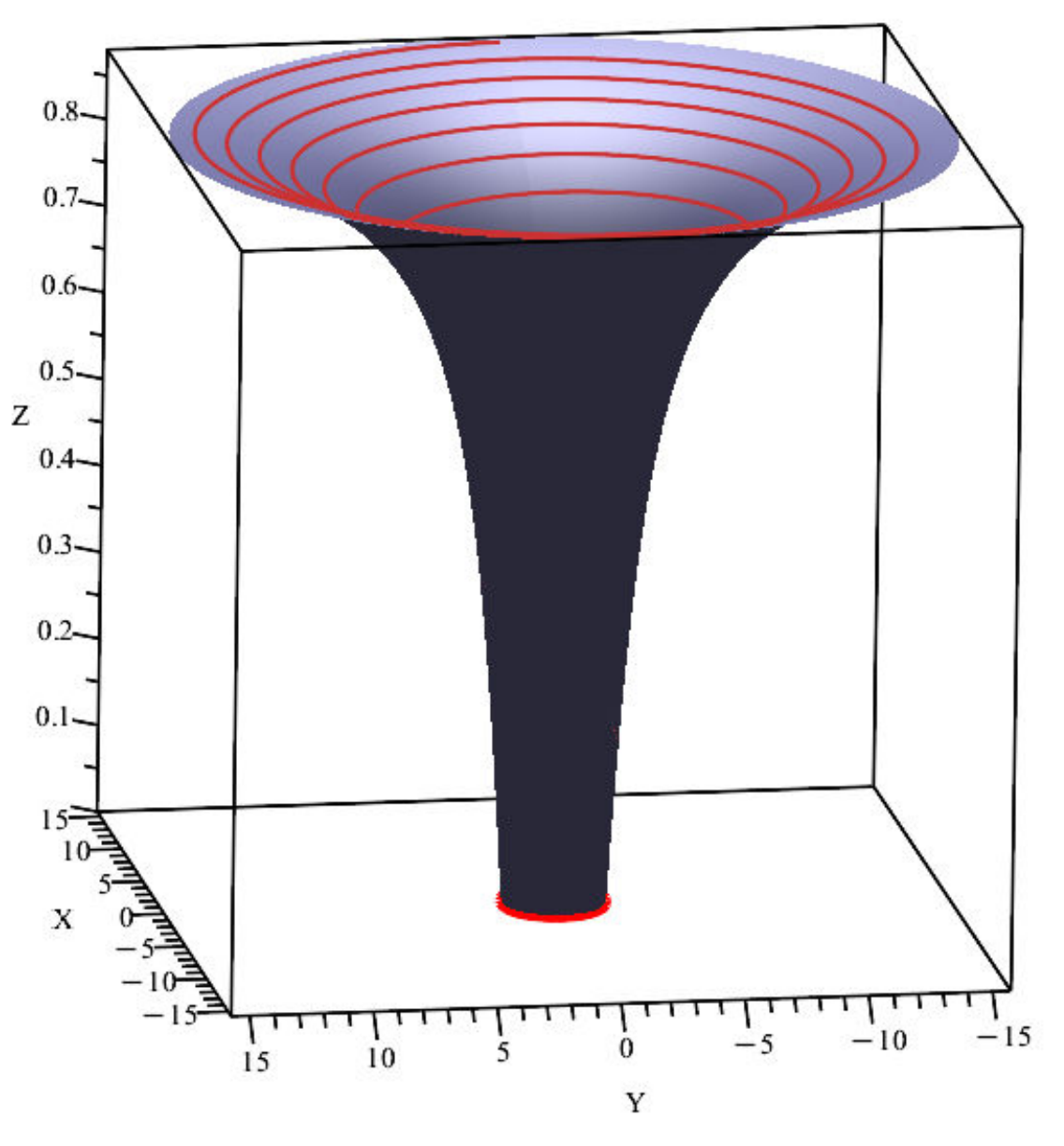}
        \subcaption{[$\zeta=2$, $M=1$, $r_h=2.0$] \newline Non-extremal BH with two horizons.}
        \label{fig:zeta0.5}
    \end{minipage}
    \begin{minipage}{0.32\textwidth}
        \centering
        \includegraphics[width=\textwidth]{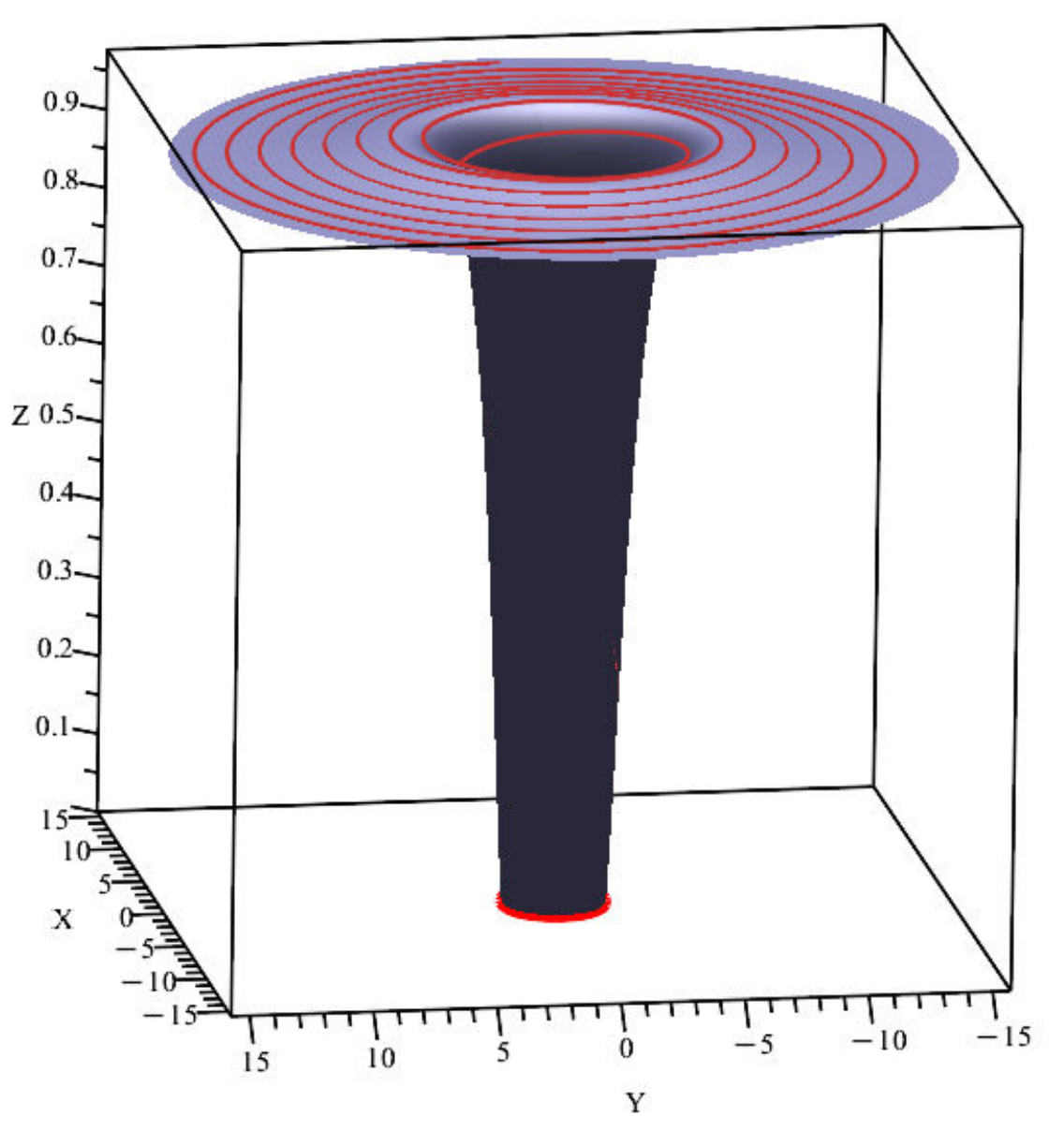}
        \subcaption{[$\zeta=5$, $M=1$, $r_h=2.0$] \newline Non-extremal BH with two horizons.}
        \label{fig:zeta2.0}
    \end{minipage}

\caption{\footnotesize 3D diagrams of the quantum-corrected BH (Model-I\&II) with metric function $g(r)$, where $M = 1$ is fixed, and $\zeta \in \{0, 2, 5\}$ varies. Each diagram features a tan surface representing the embedding from the event horizon $r_h=2.0$ to $r = 15$, a red falling trajectory, and a red thick ring at the event horizon. The transition from a single horizon at $\zeta = 0$ to two horizons for $\zeta > 0$ is accompanied by a modification of the spacetime's asymptotic flatness structure.}
    \label{fig:model1_embeddings}
\end{figure}

The geometric implications of these horizon modifications are visualized in Figure~\ref{fig:model1_embeddings}, which presents three-dimensional embedding diagrams for representative values of the quantum parameter.

Figure~\ref{fig:model1_embeddings} illustrates the dramatic evolution of spacetime geometry as quantum corrections become significant. The leftmost panel shows the classical Schwarzschild geometry ($\zeta = 0$) with its characteristic funnel-shaped embedding extending to infinity. The middle and right panels demonstrate how increasing the values of the quantum parameters ($\zeta = 2$ and $\zeta = 5$) progressively modify the geometric structure, creating more complex embeddings that reflect the configuration of two horizons.

The black falling trajectories depicted in each panel represent typical geodesic paths for infalling matter, while the red rings mark the outer horizon locations. Tan surfaces provide a visualization of the spatial curvature from the event horizon to a cutoff point at $r = 15M$. As $\zeta$ increases, the embedding surfaces show increasingly pronounced deviations from the classical Schwarzschild form, particularly in the near-horizon region, where quantum effects are strongest. 

These geometric modifications have profound implications for observable phenomena, particularly the QPO frequencies, which constitute the primary focus of this investigation. The altered spacetime structure affects particle orbits, effective potentials, and epicyclic frequencies in ways that could provide observational signatures of quantum gravity effects in astrophysical BH \cite{isz16,sec2is10}. The distinct behaviors of Model-I and Model-II, despite sharing identical horizon structures, suggest that temporal and radial metric modifications play different roles in determining observable quantities. This distinction becomes crucial when analyzing the QPO spectra and accretion dynamics that we examine in subsequent sections.

\section{QPO Dynamics from Epicyclic Motion in QCBH Spacetimes} \label{isec3}

QPOs in the X-ray flux of accreting compact objects provide one of the few direct probes of strong-field gravity in the vicinity of BH and neutron stars \cite{isz02n,isz01n}. These oscillations are commonly interpreted in terms of characteristic frequencies of particle motion in curved spacetime, such as the radial and vertical epicyclic frequencies, or combinations thereof \cite{isz03}. Although GR predicts specific relations among these frequencies in the Schwarzschild or Kerr spacetimes, several observational anomalies have motivated the study of extensions involving additional fields or modified dynamics \cite{isz04n,isz05n}.

In the QCBH concept modeling by a physical BH solution, one can analyze the motion of a neutral test particle in the vicinity of a stable circular orbit by introducing small perturbations around the equilibrium radius, denoted as $r_0$. As noted in Refs.~\cite{isz06,isz07}, these perturbations result in the emergence of epicyclic oscillations. Radial and vertical deviations exhibit linear harmonic motion, characterized by frequencies $\omega_r$ and $\omega_\theta$, respectively, while azimuthal motion maintains a steady frequency of $\omega_\phi$. The epicyclic framework serves as a valuable approach to examining the local geometry of spacetime and has been demonstrated to be effective in the analysis of nearly regular oscillations observed in accretion disks \cite{isz08n,isz09n}.

\subsection{Effective Potential and Circular Orbits}

Massive test particles moving around BHs exhibit rich and complex dynamics governed by the space-time geometry. Their motion is typically described by geodesics, which depend on the BH's gravitational potential. Circular orbits are of particular interest, especially the innermost stable circular orbit (ISCO), which marks the smallest radius at which a particle can maintain a stable orbit without spiraling into the BH. Inside the ISCO, circular orbits become unstable, causing particles either to plunge toward the event horizon or to escape outward. The properties of these orbits-such as their radius and stability-are influenced by the BH's parameters, including mass and and any modifications due to additional fields or corrections to classical gravity. Studying these orbits helps us understand accretion disk physics, gravitational wave emission, and the behavior of matter near extreme gravitational environments.

The Lagrangian density function for a static and spherically symmetric spacetime is,
\begin{eqnarray}
\mathrm{L}=\frac{1}{2}\,\Big[-f(r)\,\dot{t}^2+g(r)^{-1}\,\dot{r}^2+r^2\,\dot{\phi}^2\Big]. \label{cc2}
\end{eqnarray}

We see that the Lagrangian  is independent of $(t, \phi)$ coordinates thus the timelike geodesics admit two conserved quantities: specific energy $\mathcal{E}$ and angular momentum $\mathcal{L}$ \cite{isz10,isz11}. Focusing on the equatorial motion $(\theta = \pi/2)$, one can define
\begin{equation}
    \mathcal{E} = f(r)\,\dot t\,, 
\qquad 
\mathcal{L} = r^2\,\dot\phi\,, \label{pp1}
\end{equation}
while taking into account the normalisation condition $g_{\mu\nu}\dot x^\mu\dot x^\nu = -1$. In addition, $\dot t$ and $\dot \phi$ can also be used as eliminated quantities to derive the following quantities:
\begin{equation}
   \frac{f(r)}{g(r)}\,\dot r^2 + V_{\rm eff}(r) \;=\; \mathcal{E}^2,
    \qquad
    V_{\rm eff}(r) \equiv f(r)\!\left(1+\frac{\mathcal{L}^2}{r^2}\right),\label{pp2}
\end{equation}
where $V_{\rm eff}(r)$ encodes the gravitational attraction and centrifugal repulsion behaviors, governing the radial motion of the test particle \cite{isz12,isz13}.

\begin{figure}[ht!]
    \centering
    \includegraphics[width=0.45\linewidth]{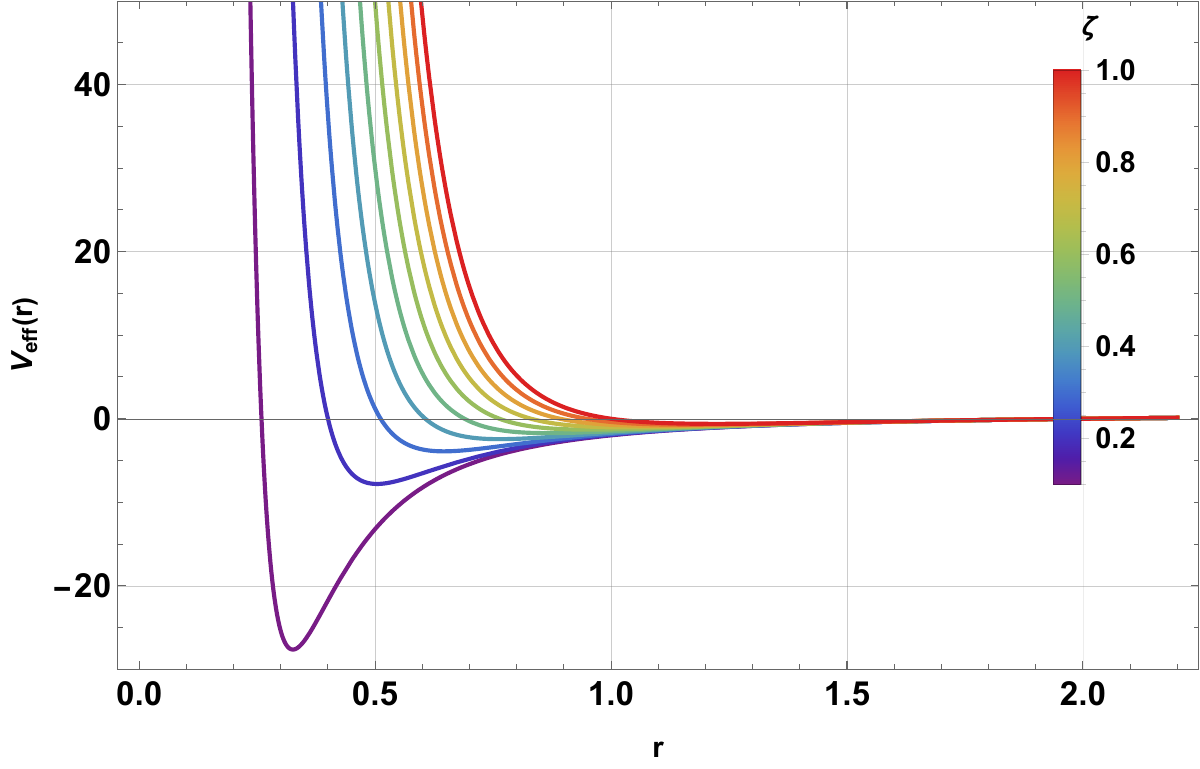}
    \caption{\footnotesize Behavior of the effective potential using the BH model-I. Here $M=1$.}
    \label{fig:potential}
\end{figure}

Circular orbits at radius $r=\mbox{const.}$, the conditions $\dot{r}=0$ and $\ddot{r}=0$ must be satisfied which implies the following relations:
\begin{equation}
    V_{\rm eff}(r)=\mathcal{E}^2,\qquad V_{\rm eff}'(r)=0.\label{pp3}
\end{equation}

Simplification of the last relation using (\ref{pp2}) gives us the specific angular momentum as,
\begin{equation}
\mathcal{L}^2(r) = \frac{r^3\,f'(r)}{2\,f(r) - r\,f'(r)} = 
\begin{cases}
\displaystyle
\frac{r^2\,\left(\dfrac{M}{r} + \dfrac{2M\,\zeta^2}{r^3}\left(1 - \dfrac{2M}{r}\right) - \dfrac{\zeta^2}{r^2}\left(1 - \dfrac{2M}{r}\right)^2\right)}
{1 - \dfrac{3M}{r} + \dfrac{2\zeta^2}{r^2}\left(1 - \dfrac{2M}{r}\right)^2 - \dfrac{2M\,\zeta^2}{r^2}\left(1 - \dfrac{2M}{r}\right)} & \text{Model-I}, \\[12pt]
\displaystyle
\frac{M\,r^2}{r - 3M} & \text{Model-II}.
\end{cases}
\label{pp4}
\end{equation}

Substituting $L^2(r)$ into the Eq. (\ref{pp3}) and after simplification results the specific energy as, 
\begin{equation}
\mathcal{E}^2(r) = \frac{2\,f(r)^2}{2\,f(r) - r\,f'(r)} =
\begin{cases}
\displaystyle
\frac{\left(1 - \dfrac{2M}{r} + \dfrac{\zeta^2}{r^2}\left(1 - \dfrac{2M}{r}\right)^2\right)^2}
{1 - \dfrac{3M}{r} + \dfrac{2\zeta^2}{r^2}\left(1 - \dfrac{2M}{r}\right)^2 - \dfrac{2M\,\zeta^2}{r^2}\left(1 - \dfrac{2M}{r}\right)} & \text{Model-I}, \\[12pt]
\displaystyle
\frac{\left(1 - \dfrac{2M}{r}\right)^2}{1 - \dfrac{3M}{r}} & \text{Model-II}.
\end{cases}
\label{pp5}
\end{equation}

\begin{figure}[ht!]
    \centering
    \includegraphics[width=0.45\linewidth]{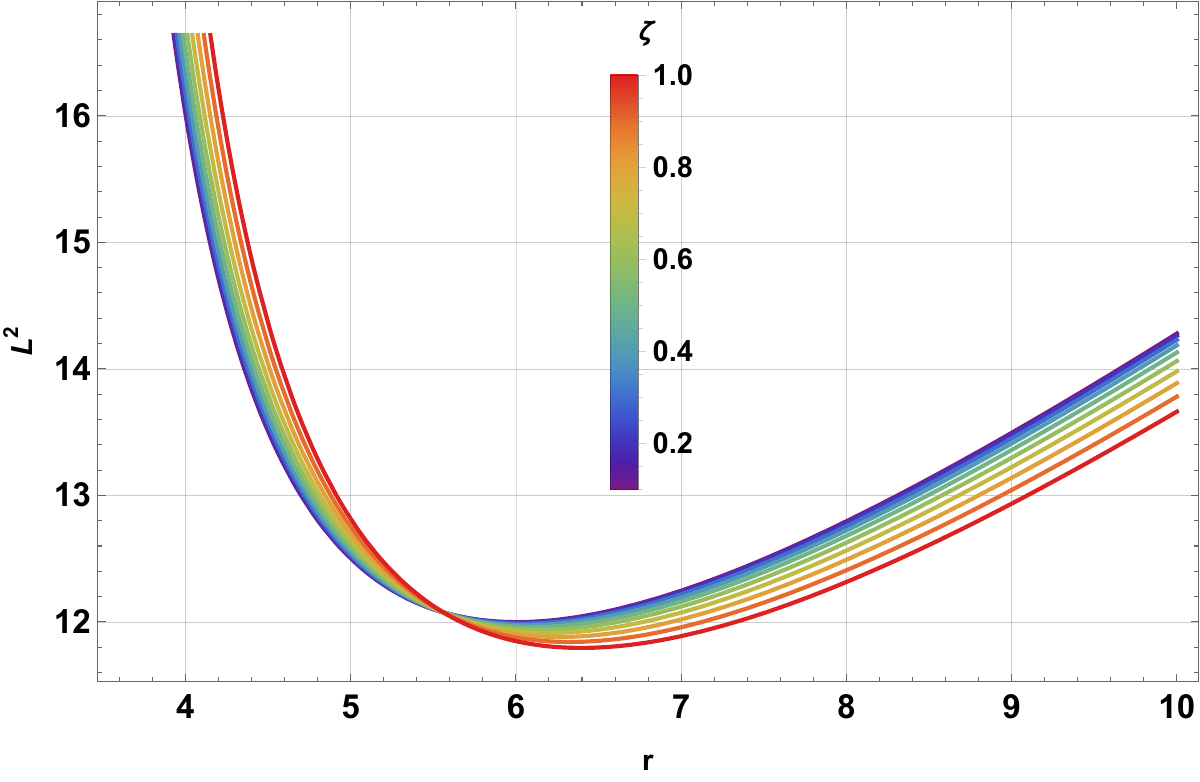}\qquad
    \includegraphics[width=0.45\linewidth]{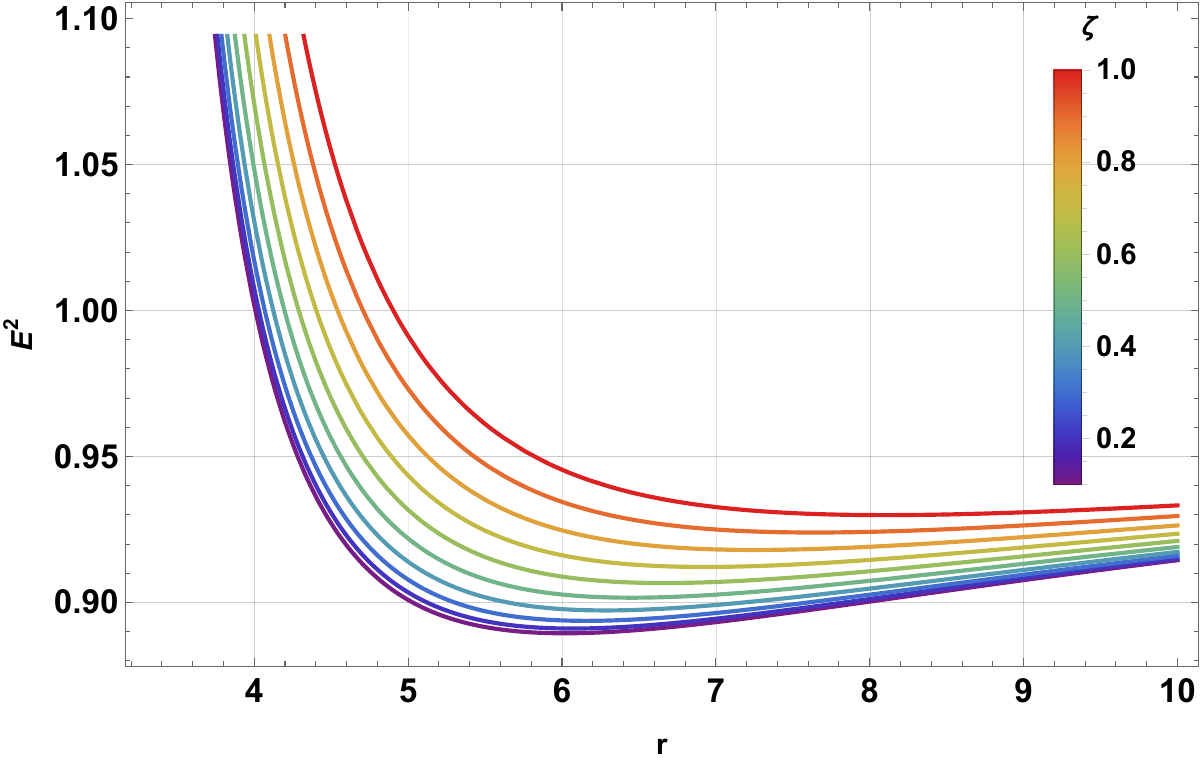}
    \caption{\footnotesize Behavior of the squared of the specific angular momentum and specific energy of test particles using BH model-I. Here $M=1$.}
    \label{fig:specific}
\end{figure}

An important aspect of the motion of massive test particles around a BH is the existence of the innermost stable circular orbit (ISCO). The ISCO represents the smallest radius at which a particle can maintain a stable circular orbit. Within this radius, circular orbits are no longer stable-any small perturbation causes the particle to either spiral inward toward the BH or escape outward. Thus, the ISCO defines the critical boundary between stable orbital motion and dynamical instability.

For a circular orbit, the following conditions must hold:
\begin{itemize}
  \item Existence of circular orbit: 
  \begin{equation}
       \frac{dV_{\text{eff}}}{dr}\Big|_{r=r_{c}} = 0.\label{pp6}
  \end{equation}
  \item Stability of circular orbit:
  \begin{equation}
    \frac{d^{2}V_{\text{eff}}}{dr^{2}}\Big|_{r=r_{c}} > 0.\label{pp7}      
  \end{equation}
  \item ISCO condition: The ISCO corresponds to the marginally stable orbit \cite{isz14}, where stability is lost, i.e.,
  \begin{equation}
      \frac{d^{2}V_{\text{eff}}}{dr^{2}}\Big|_{r=r_{\text{ISCO}}} = 0.\label{pp8}   
  \end{equation}
\end{itemize}

Using the effective potential given in Eq.~(\ref{pp2}) into the condition (\ref{pp8}) and finally using $\mathcal{L}^2(r)$ given in (\ref{pp4}) results the following relation in terms of $f(r)$ as,
\begin{equation}
    3\,f(r)\,\frac{f'(r)}{r}+f(r)\,f''(r)-2\,(f'(r))^2=0.\label{pp9}
\end{equation}
For Model-II. the ISCO radius will be $r_\text{ISCO}=6\,M$, while for the Model-I, the ISCO radius will satisfy the following polynomial relation:
\begin{align}
    &\left[1 - \dfrac{2M}{r} + \dfrac{\zeta^2}{r^2}\left(1 - \dfrac{2M}{r}\right)^2\right]\,\left[3\,\left\{\dfrac{2\,M}{r^3} + \dfrac{4\,M\,\zeta^2}{r^5}\left(1 - \dfrac{2M}{r}\right) - \dfrac{2\,\zeta^2}{r^4}\left(1 - \dfrac{2M}{r}\right)^2\right\}+\left\{-\frac{4M}{r^3} + \frac{6 \zeta^2}{r^4} - \frac{48 M \zeta^2}{r^5} + \frac{80 M^2 \zeta^2}{r^6}\right\}\right]\nonumber\\
    &-2\,\left[\dfrac{2\,M}{r^2} + \dfrac{4\,M\,\zeta^2}{r^4}\left(1 - \dfrac{2M}{r}\right) - \dfrac{2\,\zeta^2}{r^3}\left(1 - \dfrac{2M}{r}\right)^2\right]^2=0.\label{pp10}
\end{align}

Equation~(\ref{pp10}) is a polynomial in \( r \), for which obtaining an exact analytical solution is highly non-trivial. However, by applying a binomial expansion, one can derive an approximate expression for \( r \) in terms of the mass parameter \( M \) and a correction term as multiplier to \( \zeta^2 \). Alternatively, the ISCO radius can be determined numerically by choosing appropriate values for \( M \) and \( \zeta \).

\subsection{Local Epicyclic Frequencies}

To investigate the oscillatory motion of massive test particles, we consider small perturbations around stable circular orbits (SCO). When a test particle is slightly displaced from its equilibrium position in the equatorial plane, it undergoes epicyclic motion (EM), characterized by linear harmonic oscillations (HO). This approach allows for the analysis of oscillation frequencies and stability properties of the particle's trajectory.

In this setting, a test particle in a stable, circular equatorial orbit at $r_0$ may be perturbed under the following conditions:
\begin{equation}
    r(t) = r_0 + \delta r(t), 
\qquad
\theta(t) = \frac{\pi}{2} + \delta\theta(t).
\end{equation}

Thus, linearising the equations of motion allowed us to obtain two decoupled harmonic oscillator equations such that \cite{sec3is05,sec3is06}
\begin{align}
\frac{d^2\delta r}{dt^2} + \omega_r^2\,\delta r &= 0, 
\label{eq:radial-osc}\\
\frac{d^2\delta \theta}{dt^2} + \omega_\theta^2\,\delta \theta &= 0,
\label{eq:vertical-osc}
\end{align}

where, upon assessing the effective potential $V_{\rm eff}(r,\theta)$ at $\theta=\pi/2$, one finds
\begin{align}
\omega_r^2 
&= -\,\frac{1}{2\,g_{rr}}\,
   \frac{\partial^2 V_{\rm eff}}{\partial r^2}\bigg|_{\theta=\pi/2},
\label{omega-r}\\
\omega_\theta^2 
&= -\,\frac{1}{2\,g_{\theta\theta}}\,
   \frac{\partial^2 V_{\rm eff}}{\partial\theta^2}\bigg|_{\theta=\pi/2}.
\label{omega-theta}
\end{align}

Here $u^t = dt/d\tau$ is the $t$ component of the particle's four-velocity $u^\mu = dx^\mu/d\tau$.

\subsection{Frequencies Measured at Infinity}

A distant (static) observer measures coordinate-time frequencies
\begin{equation}
\Omega_i = \frac{\omega_i}{u^t}
= \frac{\omega_i}{\sqrt{\frac{2}{\,2f(r_0)-r_0 f'(r_0)\,}}},\quad i\in\{r,\theta,\phi\},
\end{equation}

since $u^t=dt/d\tau=\sqrt{\frac{2}{\,2f(r_0)-r_0 f'(r_0)\,}}$. Therefore,
\begin{align}
\Omega_\phi^2 &=- \frac{1}{2\,g_{\theta\theta}\,(u^t)^2}\,
   \frac{\partial^2 V_{\rm eff}}{\partial\theta^2}\bigg|_{r_0,\theta=\pi/2},\\
\Omega_\theta^2 &= \Omega_\phi^2,\\
\Omega_r^2 &=- \frac{1}{2\,g_{rr}\,(u^t)^2}\,
   \frac{\partial^2 V_{\rm eff}}{\partial r^2}\bigg|_{r_0,\theta=\pi/2},.
\end{align}

In physical units, the coordinate-time angular frequencies $\Omega_i$ measured by a distant observer translate into observable frequencies
\begin{equation}
    \nu_i \;=\;\frac{1}{2\pi}\,\frac{c^3}{G\,M}\,\Omega_i
[\mathrm{Hz}],
\end{equation}

where $M$ is the BH mass, and $c$ and $G$ are the speed of light and Newton's constant, respectively. Here $\Omega_r$, $\Omega_\theta$, and $\Omega_\phi$ are the dimensionless radial, latitudinal (vertical), and azimuthal angular frequencies computed in accordance with the conventions of \cite{isz15,isz16}.

Alternatively, within the parameter model $(M,\zeta)$, these frequencies can be explicitly expressed for Model-I as
\begin{align}
\label{eq:BH1_nur}
\nu_r
&=\frac{1}{2\pi}\sqrt{\frac{-2 \zeta ^4 (r-2 M)^2 \big(24 M^2-13 M r+2 r^2\big)
+3 \zeta ^2 M r^3 (r-6 M) (r-2 M)+M r^6 (r-6 M)}{r^{10}}},\\[6pt]
\label{eq:BH1_nuphi}
\nu_\theta=\nu_\phi
&=\frac{1}{2\pi}\sqrt{\frac{-8 \zeta ^2 M^2+M r^3+6 \zeta ^2 M r-\zeta ^2 r^2}{r^6}}.
\end{align}

while, for Model-II, one has
\begin{align}
\label{eq:BH2_nur}
\nu_r=\frac{1}{2\pi}\sqrt{\frac{M (r-6 M) \big(\zeta ^2 (r-2 M)+r^3\big)}{r^7}},\qquad
\nu_\theta=\nu_\phi=\frac{1}{2\pi}\sqrt{\frac{M}{r^3}}.
\end{align}

reduces\begin{align}
    \nu_\phi= \frac{1}{2\pi}\sqrt{\frac{M}{r^3}},\qquad
\nu_r = \frac{1}{2\pi}\sqrt{\frac{M(r-6M)}{r^4}},\qquad
\nu_r=\nu_\phi\sqrt{1-\tfrac{6M}{r}}
\end{align}

and the Schwarzschild ISCO appears at $r_{\rm ISCO}=6M$.

For Model-II the vertical/azimuthal frequency is exactly Keplerian and does not depend on $\zeta$. The radial frequency factorizes as
\begin{equation}
(2\pi\nu_r)^2=\frac{M (r-6 M)}{r^7}\Big(\zeta ^2 (r-2 M)+r^3\Big).
\end{equation}

Since the factor $\zeta ^2 (r-2 M)+r^3$ is positive for all physical $r>2M$ and real $\zeta$, the sign of $\nu_r^2$ is entirely controlled by $(r-6M)$. Therefore the ISCO remains exactly at $r_{\rm ISCO}^{\rm (II)}=6M$, independently of $\zeta$. For $r>6M$ the radial epicyclic frequency squared is positive (stable circular orbits), and for $r<6M$ it becomes negative (radial instability), just as in Schwarzschild.

In Model-I the radial-frequency numerator is a polynomial in $r$ containing $\zeta^2$ and $\zeta^4$ contributions (see \eqref{eq:BH1_nur}). Consequently, the zero of $\nu_r$ (the ISCO) is shifted away from $6M$ when $\zeta\neq0$. A perturbative expansion for small $\zeta$ gives the leading ISCO shift. Expanding the numerator of $(2\pi\nu_r)^2$ about $r=6M$, one finds that the linear contribution in $\zeta^2$ vanishes at $r=6M$ and the first non-zero shift appears at order $\zeta^4$. The leading correction to the ISCO is
\begin{equation}
r_{\rm ISCO}^{\rm(I)} = 6M + \frac{\zeta^4}{81\,M^3} + {\cal O}(\zeta^6).
\end{equation}

Thus, Model-I shifts the ISCO slightly outwards by a positive amount proportional to $\zeta^4$ for small $\zeta$. The shift is very small for $|\zeta|\ll M^{3/2}$.

For analytical intuition, we give the first corrections to the orbital (azimuthal) frequency in Model-I. From \eqref{eq:BH1_nuphi}:
\begin{equation}
(2\pi\nu_\phi)^2
=\frac{M}{r^3} + \zeta^2\frac{6Mr - r^2 - 8M^2}{r^6} + {\cal O}(\zeta^4).
\end{equation}

Hence, for $|\zeta|$ small,
\begin{equation}
\nu_\phi \simeq \frac{1}{2\pi}\sqrt{\frac{M}{r^3}}\left[1 + \frac{\zeta^2}{2}\,\frac{6Mr - r^2 - 8M^2}{M r^3} + {\cal O}(\zeta^4)\right].
\end{equation}

The radial frequency for Model-I also admits a small $\zeta$ expansion; at leading orders, one recovers the Schwarzschild form plus ${\cal O}(\zeta^2)$ and ${\cal O}(\zeta^4)$ corrections.

It is important to emphasize that the frequencies remain real only when the corresponding radicands are nonnegative. In Model-II the azimuthal/vertical frequency $\nu_\phi=\nu_\theta$ is manifestly real for all $r>0$, while in Model-I one must impose the condition 
\begin{equation}
-8\zeta^2 M^2 + M r^3 + 6\zeta^2 M r - \zeta^2 r^2 \ge 0,
\end{equation}

which is automatically satisfied in the asymptotic regime $(r\gg M)$ and throughout the physically relevant region outside the ISCO for typical values of the quantum parameter $\zeta$; for small $\zeta$ this merely represents a perturbative deviation from the standard Schwarzschild behavior $\propto M/r^3$. Concerning stability, the requirement $\nu_r^2>0$ ensures that circular orbits are radially stable. In Model-II one finds $\nu_r^2\propto (r-6M)$, so that stability is guaranteed only for $r>6M$, exactly as in Schwarzschild spacetime. In contrast, for Model-I the sign of $\nu_r^2$ changes at a unique physical root that defines the ISCO, located at $r_{\rm ISCO}^{\rm(I)}>2M$, which coincides with $6M$ only in the Schwarzschild limit $\zeta\to0$.

As a result, Model-II keeps the Keplerian (azimuthal/vertical) frequency identical to Schwarzschild; only the radial frequency acquires a $\zeta$-dependent amplitude. Therefore, QPO frequency ratios involving $\nu_\phi$ (e.g., $\nu_\phi/\nu_r$) will be modified chiefly by the radial-mode change. On the other hand, Model-I modifies both orbital and radial frequencies; however, the vertical / azimuthal frequency remains degenerate, $\nu_\theta=\nu_\phi$. The ISCO in Model-I is shifted outward by a small amount $\Delta r_{\rm ISCO}\sim \zeta^4/(81M^3)$ for $|\zeta|\ll M^{3/2}$, which can have a small but potentially measurable effect on the highest-frequency QPOs for compact objects if $\zeta$ is large enough.

\begin{figure}[ht!]
\centering
\includegraphics[width=8cm,height=6.5cm]{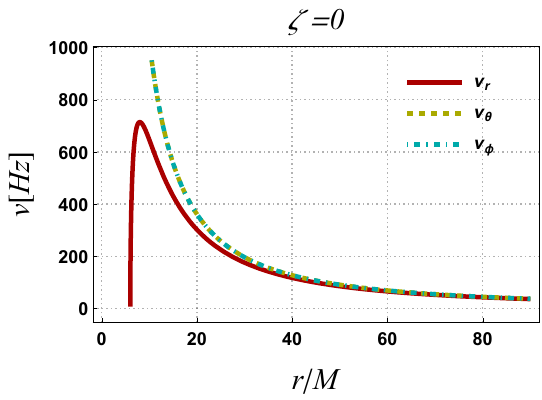}~~~~ \includegraphics[width=8cm,height=6.5cm]{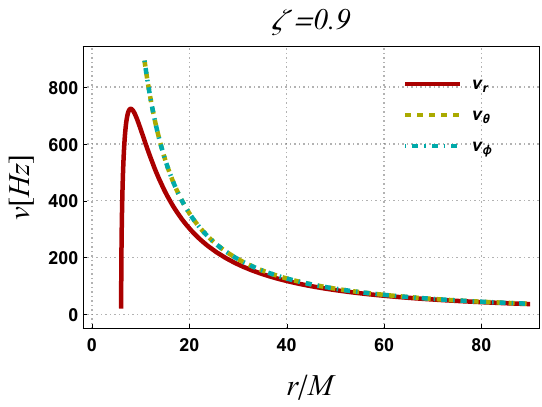}\\
\includegraphics[width=8cm,height=6.5cm]{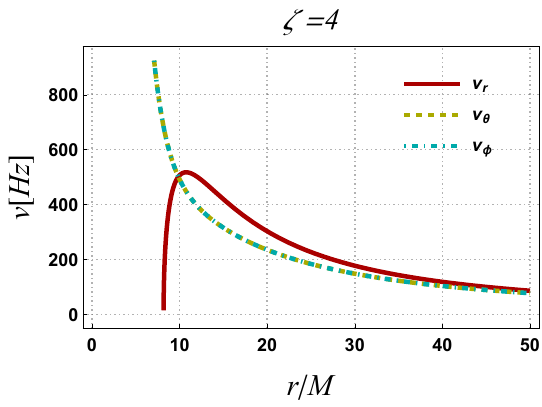}~~~~ \includegraphics[width=8cm,height=6.5cm]{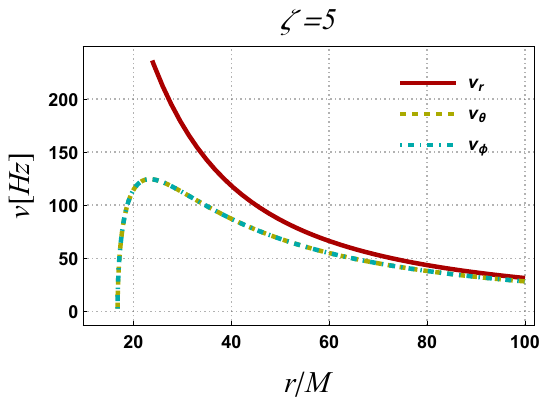}
\caption{\footnotesize Plots for the Model-I model showing the radial, vertical and azimuthal frequencies $\nu_{r,\theta,\phi}$ against $r/M$ for various values of the parameter $\zeta$ with $M=10 M_\odot$.} \label{f1}
\end{figure}

Figure \ref{f1} displays the radial, vertical and azimuthal frequencies $\nu_r,\nu_\theta,\nu_\phi$ as functions of $r/M$ for several values of the quantum parameter $\zeta$ (with $M=10M_\odot$). The plot confirms two analytic facts: (i) the vertical/azimuthal frequency remains degenerate, $\nu_\theta=\nu_\phi$, as follows from the form of $f(r)$ in Model-I, and (ii) the radial frequency departs from the Schwarzschild profile through both $\mathcal{O}(\zeta^2)$ and $\mathcal{O}(\zeta^4)$ terms; consequently the zero of $\nu_r$ (the ISCO) is shifted outward only at $\mathcal{O}(\zeta^4)$ so that for small $|\zeta|$ the overall profile remains close to Schwarzschild but with a small outward displacement of the radius where $\nu_r\to0$. The figure illustrates these effects: for small $\zeta$ the curves nearly overlap the Schwarzschild case, while for larger $\zeta$ the radial-frequency curve is visibly modified and its zero moves to larger $r$, in agreement with the perturbative result $r_{\rm ISCO}^{(I)}=6M+\zeta^4/(81M^3)+\mathcal{O}(\zeta^6)$.

The four panels in Figure \ref{f1} show the evolution of epicyclic frequencies across different $\zeta$ values: $\zeta=0$ (Schwarzschild), $\zeta=0.9$, $\zeta=4$, and $\zeta=5$. The most striking feature is the systematic modification of the radial frequency $\nu_r$ (red curves), which shows progressive deviation from the classical Schwarzschild behavior as $\zeta$ increases. For $\zeta=0.9$, the modifications are subtle, with the radial frequency showing minor deviations primarily in the near-ISCO region. However, for larger values like $\zeta=4$ and $\zeta=5$, the changes become dramatic, with the ISCO shifting significantly outward and the overall frequency profiles being substantially altered.

The azimuthal and vertical frequencies $\nu_\phi$ and $\nu_\theta$ (shown in green and cyan) remain degenerate throughout all $\zeta$ values, confirming the theoretical prediction that Model-I preserves the degeneracy between these modes despite the quantum corrections. The frequency range also changes substantially with $\zeta$: while the Schwarzschild case shows frequencies extending to approximately 1000 Hz at small radii, the $\zeta=5$ case shows dramatically reduced maximum frequencies, reflecting the outward shift of the ISCO and the modification of the effective potential structure.

\subsection{Periastron Precession}

In the realm of QCBH spacetimes, a neutral test particle located on a stable equatorial orbit at $r_0$ experiences small radial oscillations exemplified by the epicyclic frequency $\Omega_r$ (refer to Eq.~\eqref{omega-r}). The secular advance of the orbit's periastron per coordinate-time period is determined by the difference between the azimuthal frequency $\Omega_\phi$ (as defined in Eq.~\eqref{omega-theta}) and the radial epicyclic frequency such that 
\begin{equation}
    \Omega_P \;=\;\Omega_\phi \;-\;\Omega_r\,.
\end{equation}

Or, in terms of the parameter space, one has for Model-I
\begin{equation}
\Omega_p=\frac{\sqrt{\frac{M}{r^3}-\frac{\zeta ^2 (r-4 M) (r-2 M)}{r^6}}-\sqrt{\frac{-2 \zeta ^4 (r-2 M)^2 \left(24 M^2-13 M r+2 r^2\right)+3 \zeta ^2 M
   r^3 (r-6 M) (r-2 M)+M r^6 (r-6 M)}{r^{10}}}}{2 \pi }.
\end{equation}

while, for Model-II, one has
\begin{equation}
\Omega_p=\frac{\sqrt{\frac{M}{r^3}}-\sqrt{\frac{M (r-6 M) \left(\zeta ^2 (r-2 M)+r^3\right)}{r^7}}}{2 \pi }.
\end{equation}

For Model-I the two square roots entering $\Omega_p$ impose the exact reality condition
\begin{equation}
M r^3-\zeta^2(r-4M)(r-2M)\ge0,
\end{equation}

so that for $r>4M$ one has the explicit upper bound $\zeta^2\le\frac{M r^3}{(r-4M)(r-2M)}$ (while for $2M<r<4M$ the azimuthal radicand is automatically positive). Expanding for $|\zeta|\ll M^{3/2}$ yields the leading corrections
\begin{equation}
\Omega_\phi=\sqrt{\frac{M}{r^3}}-\zeta^2\frac{(r-4M)(r-2M)}{2\sqrt{M}\,r^{9/2}}+\mathcal{O}(\zeta^4),
\qquad
\Omega_r=\frac{\sqrt{M(r-6M)}}{r^2}+\zeta^2\frac{3\sqrt{M}\,(r-2M)\sqrt{r-6M}}{2\,r^5}+\mathcal{O}(\zeta^4),
\end{equation}

and hence
\begin{equation}
\Omega_p=\Bigg[\sqrt{\frac{M}{r^3}}-\frac{\sqrt{M(r-6M)}}{r^2}\Bigg]-\zeta^2\Bigg[\frac{(r-4M)(r-2M)}{2\sqrt{M}\,r^{9/2}}-\frac{3\sqrt{M}\,(r-2M)\sqrt{r-6M}}{2\,r^5}\Bigg]+\mathcal{O}(\zeta^4).
\end{equation}

Because the full radial radicand contains a nonfactorable $-2\zeta^4$ contribution, the ISCO is displaced only at order $\zeta^4$; perturbatively one finds $r_{\rm ISCO}^{(I)}=6M+\dfrac{\zeta^4}{81M^3}+\mathcal{O}(\zeta^6)$. For Model-II the radial radicand factorizes exactly,
\begin{equation}
(2\pi\Omega_r)^2=\frac{M(r-6M)\big(\zeta^2(r-2M)+r^3\big)}{r^7},
\end{equation}

so that $\zeta^2(r-2M)+r^3>0$ for all $r>2M$ and consequently the ISCO is fixed exactly at $r_{\rm ISCO}^{(II)}=6M$. Expanding for small $\zeta$ gives
\begin{equation}
\Omega_\phi=\sqrt{\frac{M}{r^3}},\qquad
\Omega_r=\frac{\sqrt{M(r-6M)}}{r^2}+\zeta^2\frac{\sqrt{M(r-6M)}\,(r-2M)}{2\,r^5}+\mathcal{O}(\zeta^4),
\end{equation}

hence
\begin{equation}
\Omega_p=\Bigg[\sqrt{\frac{M}{r^3}}-\frac{\sqrt{M(r-6M)}}{r^2}\Bigg]-\zeta^2\frac{\sqrt{M(r-6M)}\,(r-2M)}{2\,r^5}+\mathcal{O}(\zeta^4).
\end{equation}

As a result, both models recover the Schwarzschild limit $\Omega_p\simeq 3M^{3/2}/r^{7/2}$ at large $r$; the dominant $\zeta$-induced modifications are $\mathcal{O}(\zeta^2)$ (through $\Omega_\phi$ in Model-I and through $\Omega_r$ in Model-II), whereas the ISCO shift is parametrically smaller in Model-I ($\mathcal{O}(\zeta^4)$). These analytic facts determine the allowed $(r,\zeta)$ region, explain the sign-ordering $\Omega_\phi>\Omega_r$ in the stable domain, and justify why observable QPO-frequency shifts are controlled primarily by $\mathcal{O}(\zeta^2)$ corrections unless $\zeta$ is comparable to the gravitational scale $M^{3/2}$.

\begin{figure}[ht!]
\centering
\includegraphics[width=8.5cm,height=6.5cm]{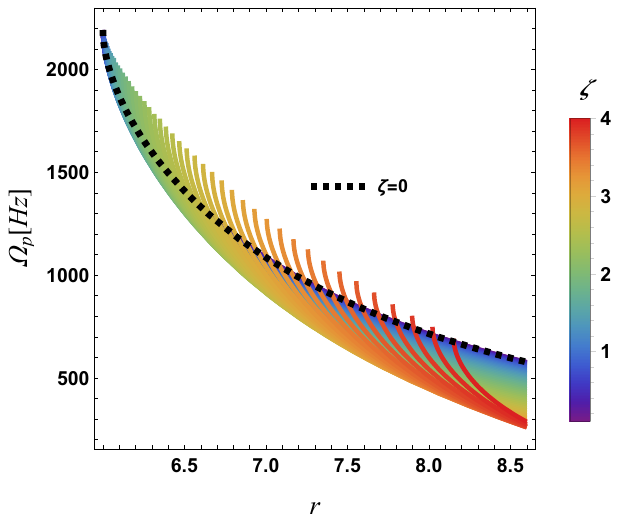}~~~~
\includegraphics[width=8.5cm,height=6.5cm]{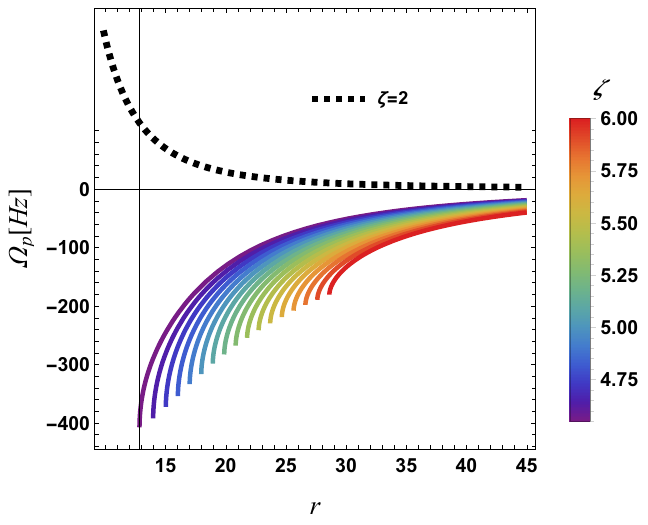}~~~~
\caption{\footnotesize Plots for the Model-I model showing the periastron frequency $\Omega_p$ for particles orbiting a BH for various values of the parameter $\zeta$ with $M=10 M_\odot$.} \label{f2}
\end{figure}

Figure \ref{f2} shows the frequency of periastron-precession $\Omega_p=\Omega_\phi-\Omega_r$ for Model-I throughout the parameter range. The plotted behavior follows directly from the exact expression: the azimuthal radicand enforces the reality condition $M r^3-\zeta^2(r-4M)(r-2M)\ge0$, which restricts the admissible $(r,\zeta)$ region and controls where $\Omega_p$ is real. In the stable domain where both radicands are nonnegative (outside the shifted ISCO), one finds $\Omega_\phi>\Omega_r$ and hence $\Omega_p>0$; the figure confirms this at large radii where all curves are asymptote to the Schwarzschild scaling $\Omega_p\simeq 3M^{3/2}/r^{7/2}$. Near the inner region, the $\zeta$-dependent corrections change the amplitude and slope of $\Omega_p$, so larger $\zeta$ produces a measurable modification of $\Omega_p$ close to the ISCO but does not qualitatively alter the positive ordering $\Omega_\phi>\Omega_r$ for physically allowed parameters.

\begin{figure}[ht!]
\centering
\includegraphics[width=8.5cm,height=6.5cm]{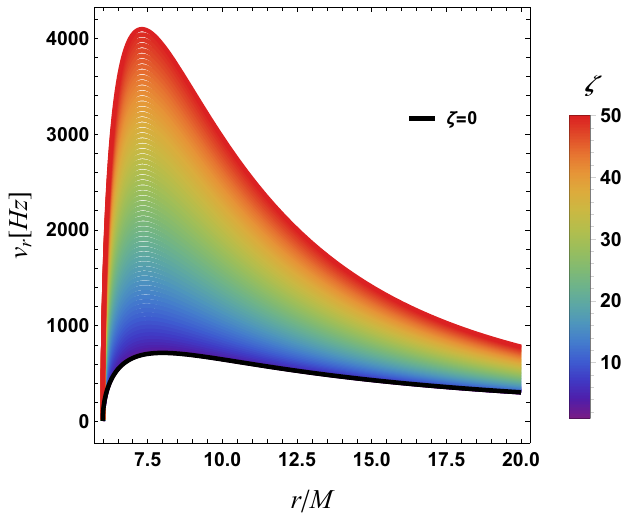}~~~~ \includegraphics[width=8.5cm,height=6.5cm]{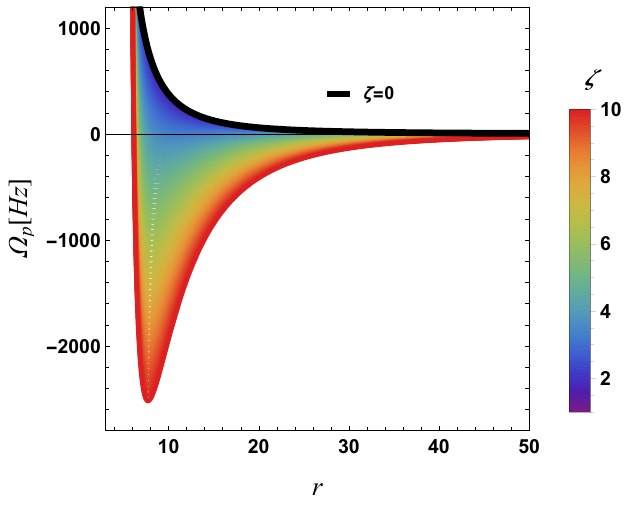}
\caption{\footnotesize Plots for the Model-II model showing the radial frequency and the periastron frequency $\Omega_p$ for particles orbiting a BH for various values of the parameter $\zeta$ with $M=10 M_\odot$.} \label{f3}
\end{figure}

Figure \ref{f3} illustrates the radial frequency and the periastron-precession frequency for Model-II. The analytic factorization in Model-II implies two immediate and exact consequences visible in the plot: (i) the azimuthal/vertical frequency remains exactly Keplerian, $\nu_\phi=\nu_\theta=(2\pi)^{-1}\sqrt{M/r^3}$, and (ii) the radial frequency factorizes as $(2\pi\nu_r)^2=M(r-6M)\big(\zeta^2(r-2M)+r^3\big)/r^7$, so the sign of $\nu_r^2$ is controlled only by $(r-6M)$ and the ISCO is fixed at $r=6M$ for all $\zeta$. Consequently, the frequency of the periastron $\Omega_p=\Omega_\phi-\Omega_r$ is monotonically reduced as $\zeta$ increases, since the $\zeta$-term increases the radial frequency radicand for fixed $r>6M$, thus increasing $\Omega_r$ and decreasing $\Omega_p$. The figure clearly exhibits these trends: $\nu_\phi$ is unchanged with $\zeta$, while $\nu_r$ and $\Omega_p$ are suppressed for larger $\zeta$ at fixed $r>6M$, and the ISCO location remains in place at $6M$, as expected analytically.


\section{BHL Accretion Simulations in QCBH Spacetimes}
\label{isec4}

With the observations of the shadows of $M87^*$ and $Sgr A^*$ BH by the EHT, the analytical formulations of alternative gravity models, solving alternative gravity metrics analytically, and describing BH shadows consistent with observations depending on the physical parameters of alternative gravity, as well as uncovering various physical phenomena such as quasi-periodic oscillations (QPOs) occurring around these BH, have emerged as an important trend \cite{isz18n,isz19n}. These groundbreaking observations have not only provided the first direct visual evidence of BH event horizons but have also opened unprecedented opportunities for testing the fundamental predictions of general relativity in the strong-field regime and exploring potential deviations that could signal new physics beyond Einstein's theory.

In this paper, we first attempt to understand the physical processes occurring around two different non-rotating quantum-corrected BH (QCBHs) by producing theoretical solutions that capture the essential physics of quantum gravitational effects. In this section, employing the Bondi-Hoyle-Lyttleton (BHL) accretion mechanism, we investigate how the dynamic structure and morphology of the shock cone formed in the plasma surrounding these BH models emerge depending on the quantum correction parameter $\zeta$ \cite{isz50,isz51,isz49}. The BHL accretion paradigm provides an ideal framework for studying matter infall in relativistic gravitational fields, as it naturally generates the shock structures and flow patterns that can trap and amplify oscillatory modes, leading to the QPO phenomena observed in X-ray binaries and other accreting compact objects.

After numerically calculating the possible QPO modes and their frequencies trapped inside the shock cone, we compare our results with both the theoretical results obtained above and with observational data from well-studied X-ray binary systems. This comprehensive approach allows us to establish direct connections between quantum gravitational modifications and potentially observable astrophysical phenomena, providing a pathway for testing quantum gravity theories through astronomical observations.

To understand the accretion mechanism formed around the two BH given in Eqs.~\ref{metric}, \ref{function-1} and \ref{function-2} through the BHL process, to reveal the dynamical structure of the resulting plasma configuration, to demonstrate the dependence of the emerging shock cone on the BH physical parameters $\zeta$, and to uncover possible quasiperiodic behaviors, we solve the general relativistic hydrodynamic equations using fixed spacetime metrics and high-resolution shock-capturing (HRSC) methods \cite{Orh1,Donmez2}. The fully relativistic solutions of these equations have been discussed in detail in \cite{Font2000LRR,Orh1}, providing the theoretical foundation for our numerical investigations.  As shown \cite{Orh4,Orh5}, by choosing the appropriate initial conditions, we numerically modeled the formation of plasma around Model-I and Model-II described in Eqs.~\eqref{metric}-\eqref{function-2} through the BHL accretion mechanism. Our computational approach employs adaptive mesh refinement techniques to resolve the complex flow structures that develop near the BH horizon and within the shock cone region. After the plasma configuration reaches a quasi-steady state, we run long-term numerical simulations and reveal that the resulting physical mechanism is a persistent process from which possible QPOs may arise. These extended simulations are crucial for capturing the full temporal evolution of instabilities and identifying the characteristic frequencies that emerge from the nonlinear coupling between hydrodynamic modes and the modified spacetime geometry.

The BHL accretion mechanism provides a natural framework for studying matter infall in the strong gravitational field regime where quantum corrections become significant \cite{isz32}. Unlike simplified analytical models that often rely on test particle approximations or linearized perturbations, our numerical approach captures the full nonlinear dynamics of relativistic fluid flow, shock formation, and the complex interplay between spacetime geometry and hydrodynamic processes \cite{isz33}.  

The computational framework incorporates several key physical processes that are essential for realistic modeling of accretion flows. These include relativistic equation of state effects, proper treatment of shock discontinuities, and careful handling of the coordinate singularities that can arise near the event horizon. The simulations are conducted using dimensionless units where the BH mass $M$ sets the fundamental scale, with our fiducial choice of $M = 10M_\odot$ providing direct connection to stellar-mass BH observed in X-ray binary systems.

\subsection{Effects of Quantum Corrections on Accretion Dynamics and Shock Cone Morphology}
\label{GRHD1}

The quantum corrections embedded in both Model-I and Model-II models fundamentally alter the spacetime geometry in ways that profoundly influence the accretion dynamics. These modifications manifest through changes in the effective potential structure, geodesic trajectories, and the overall causal structure of the spacetime \cite{isz34}. Understanding these effects is crucial for interpreting the observational signatures that could distinguish QCBHs from their classical counterparts and for establishing the parameter ranges where quantum gravitational effects become observationally accessible.

The physical origin of these modifications lies in the different ways that quantum corrections enter the metric functions of the two models. Model-I represents a scenario where quantum effects modify both the temporal $f(r)$ and radial $g(r)$ metric components simultaneously, leading to coupled alterations in the energy-momentum relationships and spatial geometry. This coupling creates a rich phenomenology where time dilation effects, gravitational redshift, and spatial curvature all deviate from their classical general relativity predictions in a correlated manner.

In contrast, Model-II preserves the classical temporal metric component while introducing quantum modifications only in the radial direction. This distinction is not merely technical but reflects fundamentally different physical scenarios for how quantum gravitational effects might manifest in realistic BH spacetimes. The preservation of classical time dilation relationships in Model-II ensures that certain aspects of the physics, particularly those related to energy conservation and asymptotic boundary conditions, remain close to their general relativity counterparts even as spatial geometry becomes significantly modified.

\subsubsection{Evolution of Shock Cone Dynamics in Model-I}
\label{GRHD2}

The accretion of matter toward the BH not only informs us about the physical properties of the surrounding phenomena, but also shows us the possible oscillatory behavior of the system. Therefore, comparing the mass accretion rate produced by matter falling toward the BH via the BHL mechanism in the Model-I QCBH with that of the Schwarzschild case provides a test of gravity in the strong-field regime and offers clues as to how Model-I can be distinguished observationally. This comparison is particularly important because accretion rate measurements represent one of the most direct observational probes of near-horizon physics, potentially accessible through X-ray flux monitoring and spectral analysis of accreting BH systems.

\begin{figure}[ht!]
\centering
\includegraphics[width=16.0cm,height=10.0cm]{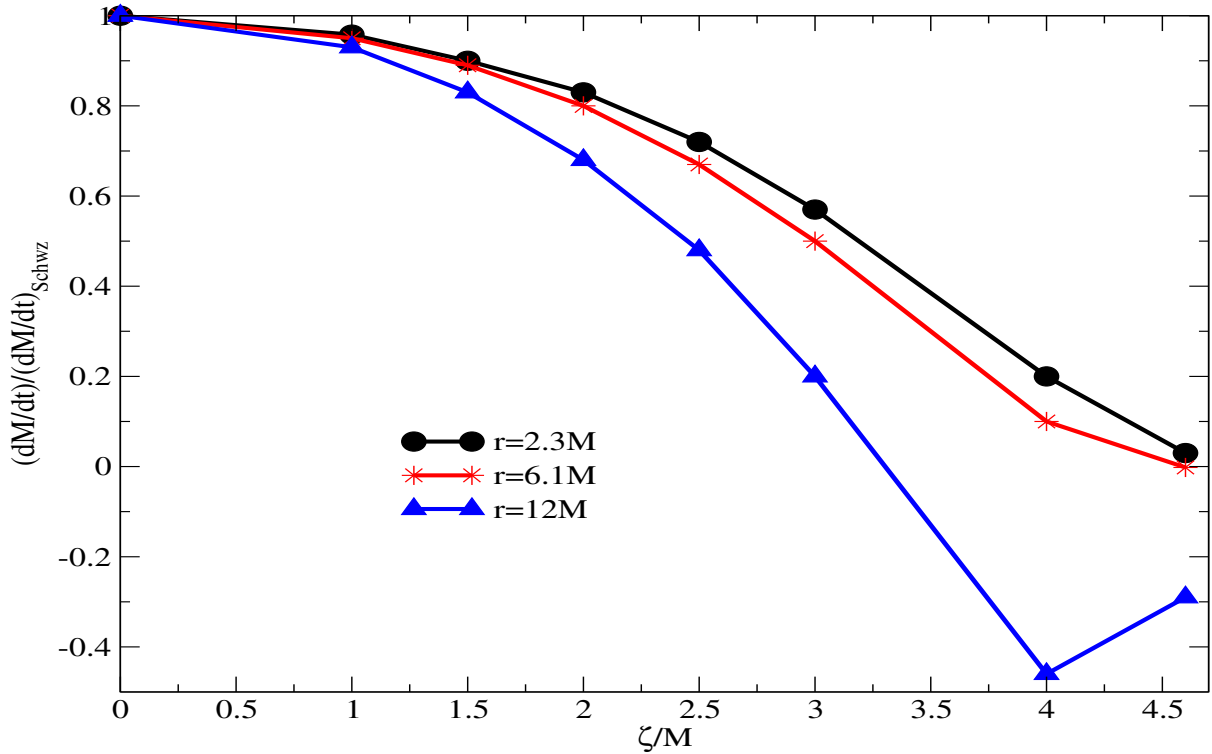}
   \caption{\footnotesize In the Model-I case, the variation of the averaged mass accretion rate, normalized by the averaged mass accretion rate around the Schwarzschild BH, is shown as a function of $\zeta$. The normalized mass accretion rate decreases with increasing $\zeta$. The rate of this decrease becomes more significant as the distance from the horizon increases. Since the stagnation point moves closer to the BH with increasing $\zeta$, the reduction in the amount of matter falling into the BH is more pronounced at $r=12M$. At the highest value of $\zeta$, compared to the Schwarzschild case, the mass accretion rate decreases by almost 95$\%$, and it even shows that at $r=12M$, instead of falling into the BH, the matter begins to move away from it.
   }  
\label{BH1_Norm_Avg_Acc}
\end{figure}

In this context, the behavior of the averaged mass accretion rate as a function of the quantum parameter $\zeta$ at different radial locations is shown in Fig.~\ref{BH1_Norm_Avg_Acc}. The choice of radial locations ($r = 2.3M$, $r = 6.1M$, and $r = 12M$) is motivated by their physical significance: the innermost location is just outside the event horizon where relativistic effects are strongest, the middle location is near the innermost stable circular orbit where orbital dynamics become unstable, and the outermost location represents the regime where Newtonian gravity begins to provide a reasonable approximation.

As can be seen from Fig.~\ref{BH1_Norm_Avg_Acc}, at all radii the mass accretion rate decreases monotonically compared to the Schwarzschild BH. This decrease proceeds regularly up to $\zeta=4M$ and beyond that value the system exhibits unstable, and hence non-physical, behavior. The emergence of this instability threshold provides an important constraint on the allowed parameter space for Model-I, suggesting that quantum corrections cannot be arbitrarily large without fundamentally disrupting the accretion process. Moreover, at $r=12M$ the mass accretion rate becomes non-physical already around $\zeta \simeq 3.2M$. The matter is effectively driven outward rather than inward, so accretion does not occur. In other words, as $\zeta$ increases, the geodesic channels that connect the outer flow to the BH become narrower and the inner edge of the efficient accretion zone is pushed outward.

This behavior reflects a fundamental modification of the effective potential structure that governs particle motion in the quantum-corrected spacetime. The narrowing of geodesic channels can be understood as a consequence of the altered curvature that deflects infalling matter away from direct radial trajectories, forcing it into increasingly complex orbital patterns that may ultimately lead to ejection rather than accretion. This phenomenon has important implications for the feeding mechanisms of BH and could potentially explain some of the observed diversity in accretion efficiency among different BH systems.

The physical interpretation of these results reveals several important aspects of quantum-corrected gravity effects. Compared to Schwarzschild, the suppression of the accretion rate is weakest very close to the horizon ($r=2.3M$), moderate near the ISCO ($r=6.1M$), and strongest farther out ($r=12M$). Physically, this means that increasing $\zeta$ in Model-I reduces the amount of matter that can accrete, with the effect becoming more pronounced at larger distances from the BH. These trends are consistent with the analytic results given in \cite{isz49,isz50,isz51,QPOmetric}. In Model-I, as $\zeta$ increases, the ISCO moves outward while the shadow radius and photon ring decreases. Both effects reduce the effective capture cross-section and the ease with which material spirals inward, thereby suppressing the mass accretion rate.

The radial dependence of the suppression effect provides important insights into the scale over which quantum corrections become significant. The fact that the effect is weakest near the horizon suggests that quantum modifications primarily influence the outer regions of the accretion flow, potentially making them more easily observable through studies of extended emission regions rather than requiring observations of the immediate near-horizon environment.

The three curves in Fig.~\ref{BH1_Norm_Avg_Acc} corresponding to different radial locations ($r=2.3M$, $r=6.1M$, and $r=12M$) demonstrate the radial dependence of quantum corrections. The curve at $r=2.3M$ (black circles) shows the most gradual decline, maintaining approximately 60$\%$ of the Schwarzschild accretion rate even at $\zeta=4M$. This behavior reflects the fact that very close to the horizon, the dominant gravitational effects still resemble those of classical GR, with quantum corrections providing only secondary modifications. The persistence of relatively high accretion rates in the near-horizon region suggests that the innermost parts of accretion disks may remain largely unaffected by quantum modifications, preserving many of the classical predictions for emission from these regions.

In contrast, the curve at $r=12M$ (blue triangles) exhibits the most dramatic suppression, with the accretion rate dropping to nearly zero at $\zeta \approx 3.2M$ and actually becoming negative for higher $\zeta$ values. This negative accretion rate indicates that matter is being expelled from the BH rather than accreted, representing a fundamental breakdown of the classical accretion paradigm. The physical mechanism underlying this expulsion involves the modification of effective potentials in such a way that the gravitational attraction is overcome by quantum-induced repulsive effects, creating an effective barrier to infall at large radii.

The intermediate behavior at $r=6.1M$ (red crosses) provides a bridge between these two extremes, showing significant suppression but without the complete reversal of flow direction observed at larger radii. This intermediate regime is particularly interesting from an observational perspective, as it represents a parameter range where quantum effects are substantial but not so extreme as to completely disrupt accretion, potentially providing the optimal conditions for detecting quantum signatures in realistic astrophysical systems. On the other hand, the BHL mechanism generates a shock cone around all BH defined in GR and alternative gravity theories. Understanding the dynamical structure of this cone not only allows gravity to be tested in the strong-field region but also significantly affects the excitation of fundamental modes trapped within the shock cone. The shock cone represents a quasi-static structure that forms when supersonic matter flow encounters the gravitational field of the BH, creating a region of compressed, heated material that can support various types of oscillatory modes.

\begin{figure*}[htp!]
\centering
\includegraphics[width=16.0cm,height=10.0cm]{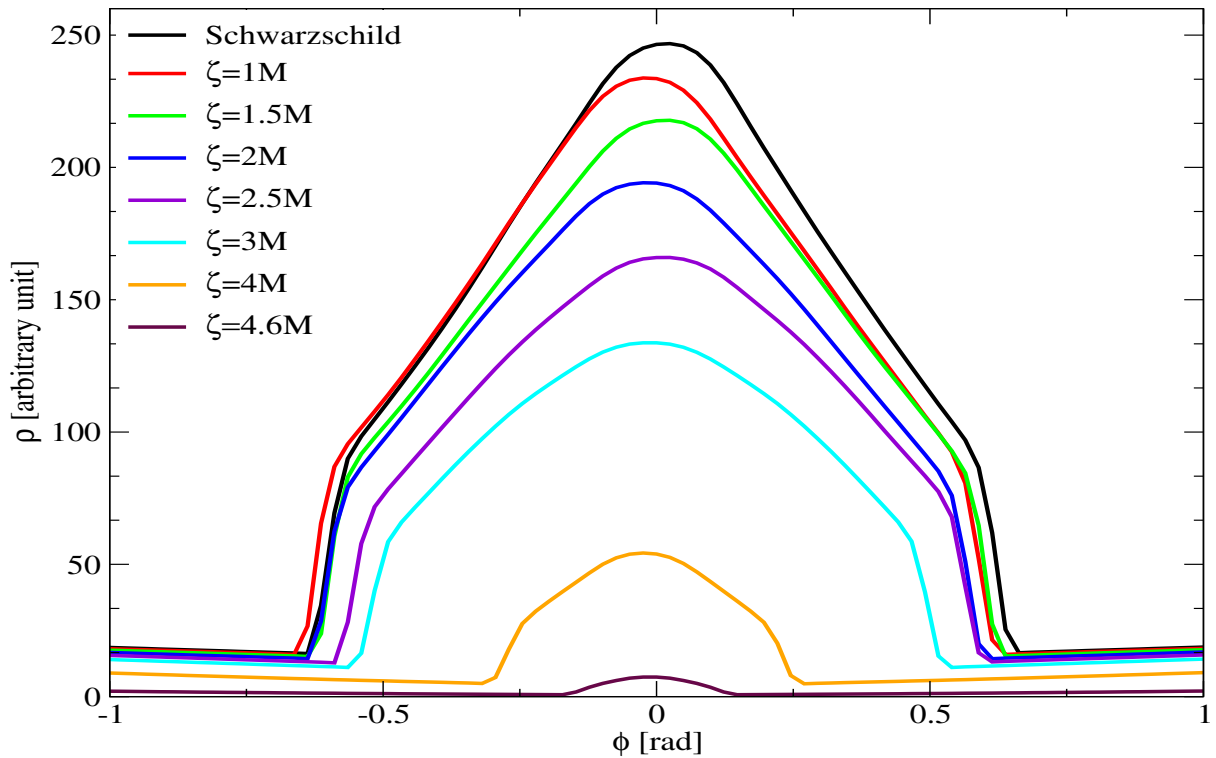} 
    \caption{\footnotesize At $r=2.66M$, the variation of the rest-mass density in the azimuthal direction has been shown for different $\zeta$ values in both the Schwarzschild and Model-I cases. It has been observed that as $\zeta$ increases, the density of the matter accreted inside the shock cone formed around the BH decreases. This decrease is particularly significant at large values of $\zeta$. Additionally, the opening angle of the shock cone decreases with increasing $\zeta$. These changes can lead to variations in the QPOs formed around BH.
   }
\label{BH1_denCut}
\end{figure*}

In Fig.~\ref{BH1_denCut}, at $r=2.66M$, very close to the event horizon, the azimuthal variation of the rest-mass density is shown, revealing the structure of the shock cone formed around the BH in the Model-I case and demonstrating how the cone dynamical structure changes with $\zeta$. The choice of this radial location is motivated by the desire to probe the regime where both gravitational and quantum effects are strongest, providing the most sensitive test of the quantum correction predictions.

Physically, Fig.~\ref{BH1_denCut} tells us how matter accretes onto the BH, how it distributes around it, and, in particular, how it forms in the presence of a shock cone created by supersonic inflow and accretion dynamics. For the Schwarzschild case ($\zeta=0M$), the amplitude of the shock cone density is maximal. As expected, since the BH does not rotate, the maximum amplitude occurs almost exactly at $\phi=0$ rad, reflecting the spherical symmetry of the underlying spacetime. The sharp, well-defined peak structure indicates efficient shock formation and compression of the infalling material.

As $\zeta$ increases, the peak density gradually decreases. At $\zeta=1M$ and $\zeta=1.5M$, the maximum amplitude remains comparable to the Schwarzschild case, but for $\zeta \geq 3M$ the maximum density of the cone is significantly weakened. At $\zeta=4.6M$, the shock cone is almost in the stage of disappearing. This systematic evolution demonstrates the progressive weakening of shock formation as quantum corrections become stronger, reflecting the fundamental changes in the underlying spacetime geometry that support shock propagation.

The density profiles shown in Fig.~\ref{BH1_denCut} reveal several key features of the quantum-modified accretion process. The Schwarzschild profile (black line) exhibits a sharp, well-defined peak at $\phi=0$ with a characteristic width of approximately 1 radian, representing the classical shock cone structure. This profile reflects the efficient focusing of matter along the axis of symmetry and the strong compression that occurs when supersonic flow encounters the gravitational field.

As $\zeta$ increases through the sequence $\zeta=1M$ (red), $\zeta=1.5M$ (green), $\zeta=2M$ (blue), the peak density systematically decreases while the cone width remains relatively unchanged. However, for larger quantum corrections ($\zeta=2.5M$, $\zeta=3M$, $\zeta=4M$), both the peak density and the cone opening angle are significantly reduced. This dual evolution—decreasing both amplitude and width—indicates that quantum corrections affect both the efficiency of matter compression and the geometric focusing properties of the gravitational field.

The evolution of the cone opening angle provides particularly important information about the modification of geodesic focusing properties. In classical general relativity, the opening angle is determined by the balance between gravitational attraction and centrifugal effects, with the specific value depending on the details of the spacetime geometry. The systematic decrease in opening angle with increasing $\zeta$ suggests that quantum corrections modify this balance in a way that favors tighter focusing, potentially leading to more collimated flows and different emission patterns.

This behavior of the rest-mass density clearly shows that $\zeta$ strongly affects matter accretion, weakening the compression and shock formation. This, in turn, demonstrates that $\zeta$ seriously modifies the spacetime itself, as we previously discussed, which means that instead of matter falling supersonically directly into the BH, it tends to spiral around and is expelled beyond the computational domain. Since this would manifest observationally as a decrease in mass accumulation around BH sources, such a feature provides an important distinguishing signature for Model-I.

The implications for observational astronomy are significant. The weakening of shock formation would lead to reduced X-ray emission from shock-heated material, potentially observable as changes in the hard X-ray spectrum of accreting BH. Similarly, the modification of flow patterns could affect the structure of relativistic jets, which are thought to be closely connected to accretion disk dynamics. These effects provide potential observational targets for testing Model-I predictions using current and future X-ray missions.

Understanding the physical phenomena that occur around BH in a strong gravitational field and the strength of the shock waves that form can be achieved by revealing the physical properties of the accreting matter and the shock mechanism itself. We illustrated this earlier by plotting the behavior of different parameters in Figs.~\ref{BH1_denCut} and \ref{BH1_Norm_Avg_Acc}. Here, we present the radial and azimuthal velocities of the matter. From these, $v_r/c$ tells us where and how strongly the flow plunges and shocks, whereas $v_{\phi}/c$ tells us how the flow spins and transports angular momentum. By examining these behaviors, one can infer the shock cone opening angle, the shock strength, compression or heating, the mass and angular-momentum fluxes, turbulence, the sonic structure, possible instabilities, and the expected brightness characteristics.

The velocity field analysis provides crucial information about the momentum transport and energy dissipation mechanisms that operate in the quantum-corrected spacetime. Radial velocity patterns reveal the efficiency of matter infall and the locations where kinetic energy is converted to thermal energy through shock heating. Azimuthal velocity patterns indicate the degree of angular momentum transport and the strength of rotational motion that could drive jet formation or disk instabilities.

\begin{figure*}[!htp]
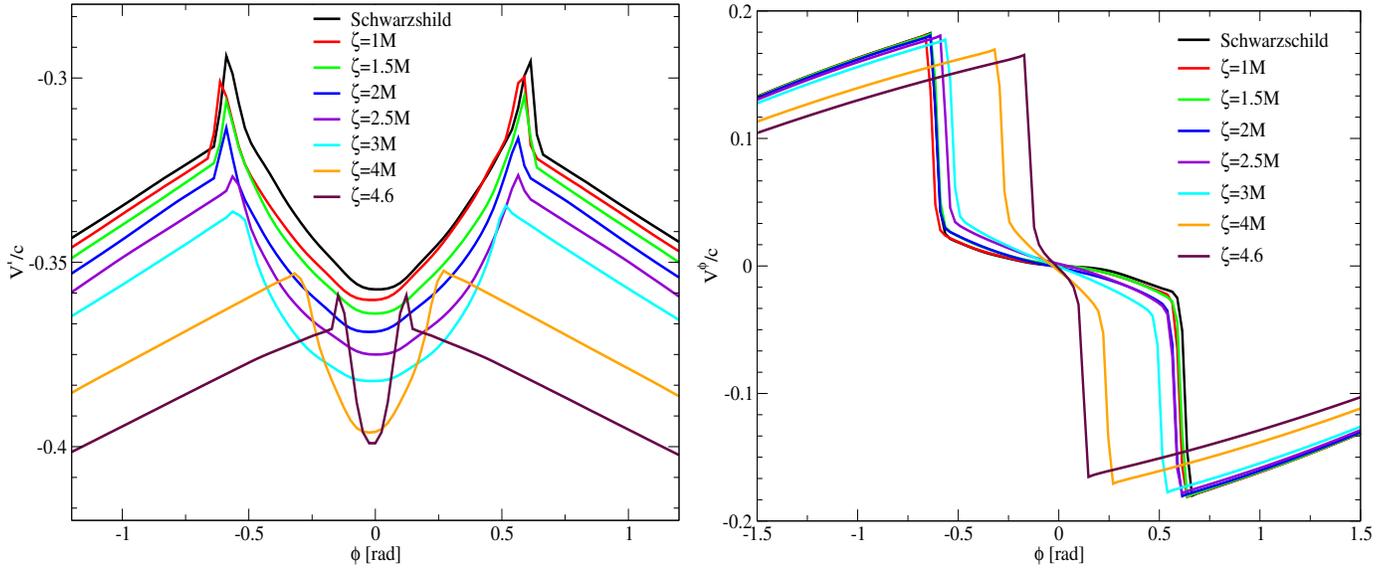

\centering
\includegraphics[width=9.0cm,height=7.5cm]{BH1radial_veloc_fixed_r.eps} 
\includegraphics[width=9.0cm,height=7.5cm]{BH1azimuthal_veloc_fixed_r.eps} 
    \caption{\footnotesize This shows the variation of matter velocities around BH in a strong gravitational field at $r=2.66M$ for different $\zeta$ values in the Schwarzschild and Model-I cases. The plot on the left shows the variation of radial velocity along the azimuthal path, while the plot on the right shows the variation of azimuthal velocity. At the shock locations of the shock cone, the radial velocity decreases, whereas the azimuthal velocity reaches its maximum in both positive and negative directions. On the other hand, matter rapidly accretes toward the BH through the interior of the cone. The velocity plots clearly show that the behavior of matter in the strong gravitational field changes with increasing $\zeta$ values.
   }
\label{BH1_VelocCut}
\end{figure*}

In the left panel of Fig.~\ref{BH1_VelocCut}, the behavior of the radial velocity at $r=2.66M$ (i.e., at the location closest to the horizon) is shown. As seen in the figure, the velocity reaches its deepest (most negative) value at $\phi=0$ rad. In the Schwarzschild case, two sharp spikes appear in the range $\phi=0.5-0.6$ rad. As $\zeta$ increases, the entire curve shifts toward more negative values, which means that the matter plunges toward the BH at higher speed. This is also a hallmark of a narrowing and hardening shock cone. Moreover, as $\zeta$ increases, the shock cone becomes progressively narrower, and photon and matter orbits are focused more tightly. Thus, while the photon impact parameter $b_{ph}$ shrinks with $\zeta$, the photon sphere radius $r_{ph}$ remains fixed at $3M$. Lensing and gravitational compression therefore increase near the axis, consistent with the deeper $v_r/c$ trough at $\phi=0$.

The evolution of the velocity spikes with increasing $\zeta$ provides important information about the shock structure and its modification by quantum effects. In the classical Schwarzschild case, the spikes represent locations where the shock front intersects the measurement radius, creating localized regions of decelerated flow. The systematic shift of these features with increasing $\zeta$ indicates that quantum corrections modify not only the overall flow pattern but also the detailed structure of shock fronts and their propagation characteristics.

At the spike locations, the radial speed is partially quenched by the shock, which raises the rest-mass density and pressure there. These two spike regions actually help accelerate the flow. Consequently, where $v_r/c$ is the least negative, the gas piles up and brightens, whereas the most negative portions along the axis mark fast and diluted streams. Altogether, the $\zeta$-dependent changes seen in the radial velocity given in left panel provide sharp physical signatures by which the Model-I QCBH can be observationally distinguished from a Schwarzschild BH.

The relationship between velocity patterns and observable emission provides a direct connection between our theoretical predictions and potential astronomical observations. Regions of gas accumulation (where $v_r/c$ is least negative) would appear as bright spots in X-ray images, while regions of fast, diluted flow would contribute less to the observed flux. The systematic evolution of these patterns with $\zeta$ suggests that Model-I could be distinguished from classical BH through careful analysis of X-ray flux distributions and their temporal variations.

In the right panel of Fig.~\ref{BH1_VelocCut}, we compute the azimuthal velocity $v_{\phi}/c$ at the same radius $r=2.66M$ where the radial velocity was evaluated. As seen in the figure, the profile exhibits a shock shear pattern: on the upstream side ($\phi<0$) there is a positive plateau and then decelerates rapidly, reaching $v_{\phi}/c =0$ near $\phi=0$. Afterward, there is a sharp and symmetric jump to a negative plateau on the downstream side ($\phi>0$). The jump locations mark the flanks of the shock cone. As $\zeta$ increases, these jump locations move toward $\phi=0$, and the plateaus grow in magnitude (larger $\lvert v_{\phi}/c \rvert$). That is, the cone narrows and the azimuthal shear strengthens.

The azimuthal velocity profile provides crucial information about angular momentum transport and the development of rotational instabilities. The sharp transitions between positive and negative plateaus indicate the presence of strong velocity shear that could drive turbulence and magnetic field amplification in magnetized flows. The systematic increase in shear strength with $\zeta$ suggests that quantum corrections could enhance these instability mechanisms, potentially leading to more efficient angular momentum transport and stronger jet production.

Physically, a larger $\zeta$ tightens the geodesic focusing of matter falling toward the BH within the shock cone, so the flow picks up more azimuthal velocity as it is deflected around the hole, and the shear layer along the cone walls becomes thinner and stronger. Consequently, in this strong gravitational region, the velocity profiles imply stronger compression and heating, together with faster, more dilute streams along the axis. In short, the evolution of both velocity components with increasing $\zeta$ demonstrates the profound impact of quantum corrections on the hydrodynamic structure of the accretion flow. The narrowing of the shock cone, the intensification of velocity shear, and the enhancement of gravitational focusing all represent potential observational signatures that could be detected through careful analysis of X-ray timing and spectral data from accreting BH.

\subsubsection{Evolution of Shock Cone Dynamics in Model-II}
\label{GRHD3}

In Section \ref{GRHD2} we discussed the case of Model-I, here we now repeat the same analysis for Model-II. The comprehensive comparison between these two models provides crucial insights into how different types of quantum corrections manifest in observable astrophysical phenomena. While Model-I demonstrated dramatic modifications to accretion dynamics, Model-II represents a fundamentally different scenario where quantum effects are more subtle but potentially longer-lasting in their impact on system behavior.

\begin{figure*}[!htp]
\centering
\includegraphics[width=16.0cm,height=10.0cm]{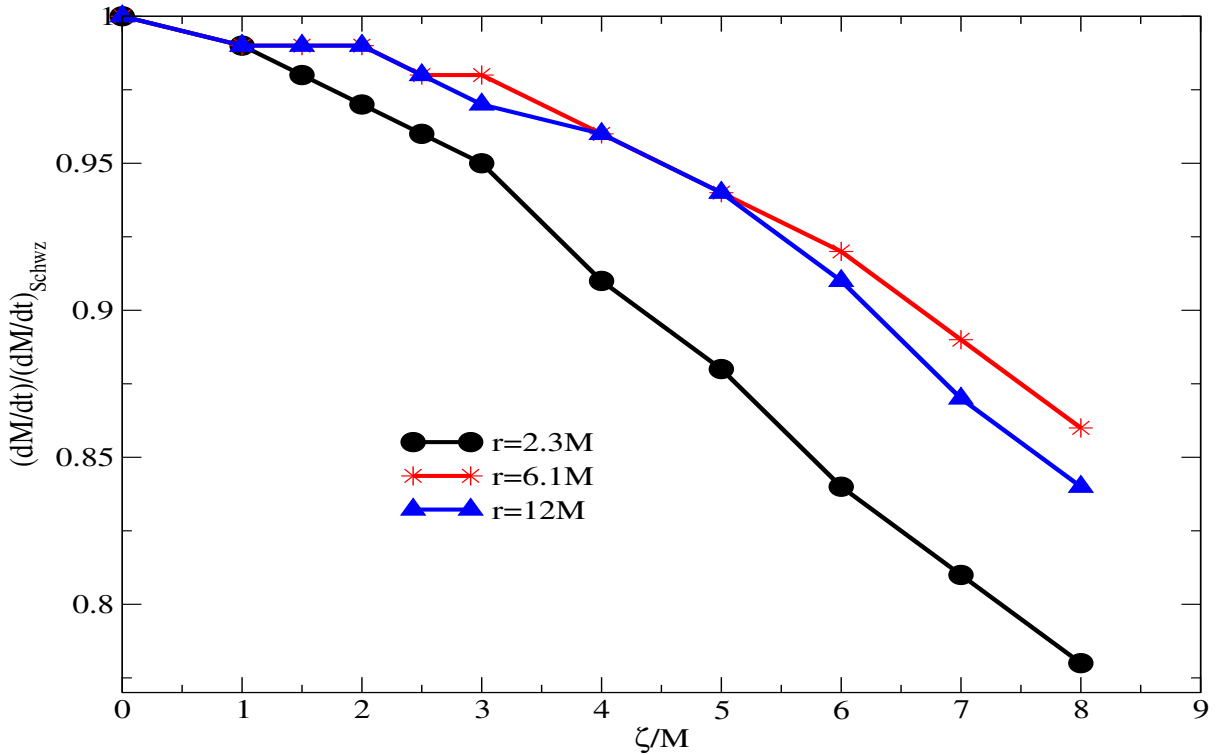} 
     \caption{\footnotesize The same analysis as in Fig.\ref{BH1_Norm_Avg_Acc} has been performed, but this time the ratios are shown for the Model-II case. Unlike the Model-I case given in Fig.\ref{BH1_Norm_Avg_Acc}, the rate of decrease is much slower and follows a more linear trend. The behavior of the decrease at each radial point is similar. When compared with the Schwarzschild case, the decrease in the mass accretion rate with increasing $\zeta$ values is at most around 20$\%$.
    }
\label{BH2_Norm_Avg_Acc}
\end{figure*}

First, in the case of Model-II, the average mass accretion rate normalized with respect to the Schwarzschild case is shown in Fig.\ref{BH2_Norm_Avg_Acc} as a function of $\zeta$. At different radial locations, the normalized mass accretion rate decreases monotonically with increasing $\zeta$. At $r=2.3M$, the reduction in the accretion rate is the strongest. For large values of $\zeta$, the accretion rate shows a tendency to decrease by about 15–20$\%$ compared to Schwarzschild. This indicates that the strong gravitational field is the region most sensitive to metric modifications because the flow is deeper in the potential well. On the other hand, at $r=6.1M$, the decrease in the accretion rate is slower compared to $r=2.3M$, showing a reduction of approximately 10$\%$ in large $\zeta$. At $r=12M$, the change exhibits a similar trend to that at $r=6.1M$. As a result, in the relatively weaker gravitational field regions at $r=6.1M$ and $r=12M$, the effect of the quantum correction term is comparatively smaller. Taking this into account, the strong-gravity regime can be tested using the metric given in Model-II.

The more modest changes observed in Model-II compared to Model-I reflect the different nature of quantum corrections in this model. By preserving the classical temporal metric component, Model-II maintains the standard relationships between energy, time dilation, and asymptotic boundary conditions. This preservation ensures that the large-scale structure of the accretion flow remains close to its classical form, even as spatial geometry undergoes significant modifications near the BH.

For Model-II, the results obtained from our numerical simulations are consistent with those reported in \cite{isz51}. As discussed in \cite{isz51}, the effect of $\zeta$ on $b_{ph}$ and the photon sphere is very small. At the same time, the brightness shows only minor variations, and the ring width exhibits a very slight shift. These findings support the results we obtain for Model-II in our numerical simulations. However, the variation that we observe in the accretion rate is somewhat more pronounced than the changes reported in \cite{isz51}. The reason is that in \cite{isz51} the analysis was based on ray-tracing, where only photon trajectories were considered, whereas in our case we perform relativistic hydrodynamical modeling and include the effects of gas dynamics, namely the forces arising from gas pressure. Therefore, it is expected that suppression dependent on the BH horizon appears more clearly in our numerical results and Fig.\ref{fig:model1_embeddings}.

This comparison highlights an important distinction between ray-tracing studies, which probe the geometric properties of spacetime through photon propagation, and hydrodynamical simulations, which capture the full dynamics of matter flow including pressure forces, shock formation, and nonlinear coupling between different physical processes. The enhanced sensitivity of hydrodynamical probes to quantum corrections suggests that accretion studies may provide more stringent tests of quantum gravity theories than purely geometric observations.

The linear trend observed in Fig.\ref{BH2_Norm_Avg_Acc} contrasts sharply with the exponential-like behavior seen in Model-I, indicating fundamentally different scaling relationships between quantum corrections and accretion efficiency. This linear dependence suggests that Model-II modifications accumulate gradually rather than exhibiting the threshold behavior characteristic of Model-I, potentially making Model-II signatures more easily detectable at smaller values of the quantum parameter.

The similar behavior at all radial points in Model-II indicates that quantum corrections affect the entire accretion flow in a coherent manner, rather than creating the strong radial gradients observed in Model-I. This coherent modification could lead to more stable, predictable changes in observable quantities, making Model-II potentially easier to distinguish from classical general relativity through systematic parameter estimation studies.

\begin{figure*}[!htp]
\centering
\includegraphics[width=16.0cm,height=10.0cm]{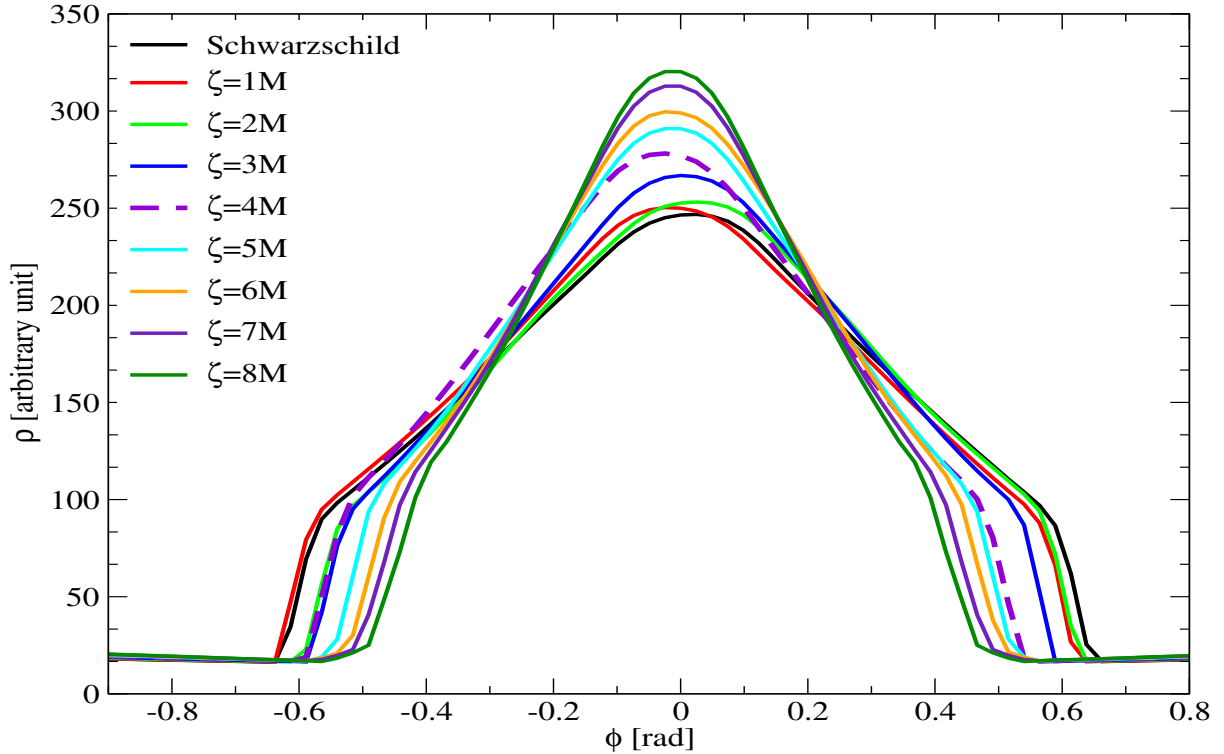} 
     \caption{\footnotesize As in Fig.\ref{BH1_denCut}, the variation of the rest-mass density is shown for the Schwarzschild and Model-II cases. Unlike the Model-I case, with increasing $\zeta$ values, the density of the matter trapped inside the shock cone increases, while the opening angle of the cone decreases. Similar to Model-I, the QPO modes trapped inside the cone are significantly affected by the change in the physical structure.
    }
\label{BH2_denCut}
\end{figure*}

The behavior of the shock cone formed around Model-II as a function of $\zeta$ in the strong gravitational field not only allows us to understand how the metric function changes in the QCBH case but also provides a comparison with the Schwarzschild BH model. In this context, the variation of the rest-mass density in the azimuthal direction at $r = 2.66M$ for different $\zeta$ values is shown in Fig.\ref{BH2_denCut}. A strong shock cone is clearly observed. For $\zeta=1M$, $2M$, and $3M$, the density profile of the cone exhibits behavior almost identical to that of the Schwarzschild case, although the peak density increases slightly. For $\zeta=4M$, $5M$, and $6M$, the peaks become sharper, indicating stronger compression of the density. At the same time, the profile also becomes more asymmetric, with a slower decline for positive $\phi$. For $\zeta=7M$ and $8M$, the peak broadens and increases further, while the opening angle of the shock cone becomes larger. This modifies the frequencies of the modes trapped inside the cone, a point that is discussed later. Therefore, $\zeta$ modifies the scattering and focusing of trajectories, allowing more matter to be collected in the same region. Compared with BH-II simulations shown in \cite{isz51}, unlike in that case, the effect of $\zeta$ in Model-II is more pronounced. This may be due to the fact that, in contrast to the setup in \cite{isz51}, the pressure force of the matter is included in the numerical modeling around the BH in Model-II.

The enhancement of shock compression in Model-II represents a particularly interesting phenomenon that contrasts with the weakening observed in Model-I. This enhancement suggests that while Model-II preserves the overall structure of the accretion flow, it modifies the detailed physics of shock formation in a way that leads to more efficient matter compression and heating. These effects may observationally appear as increased X-ray emission from shock-heated matter, serving as a potential indicator for differentiating Model-II from both classical BH and Model-I systems.

The contrasting behavior between Model-I and Model-II in terms of density evolution provides one of the clearest discriminants between these quantum correction scenarios. While Model-I shows systematic density suppression with increasing $\zeta$, Model-II demonstrates density enhancement, suggesting fundamentally different mechanisms by which quantum corrections influence matter compression and shock formation. This difference has profound implications for observable signatures, as enhanced density would lead to brighter X-ray emission from shock-heated material, while density suppression would have the opposite effect. The decrease in opening angle observed in Model-II, combined with density enhancement, suggests that quantum corrections in this model lead to more efficient gravitational focusing without the instabilities that plague Model-I at large $\zeta$ values. This stable focusing behavior could make Model-II systems particularly effective at producing collimated outflows and jets.

\begin{figure*}[!htp]
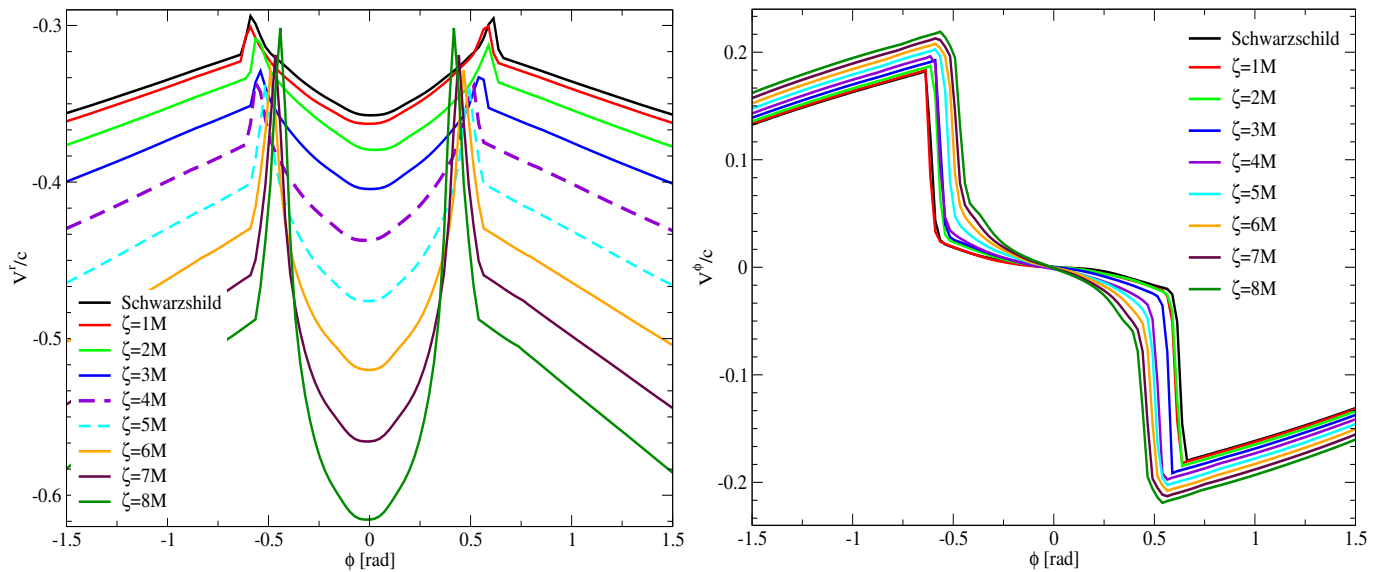

\centering
\includegraphics[width=9.0cm,height=7.5cm]{ BH2radial_veloc_fixed_r.eps} 
\includegraphics[width=9.0cm,height=7.5cm]{ BH2azimuthal_veloc_fixed_r.eps} 
     \caption{\footnotesize This is the same as Fig.\ref{BH1_VelocCut}, but this time the azimuthal variations of radial and azimuthal velocities are shown for the Model-II case. Although the behavior differs from Model-I, it is clearly seen that these velocities undergo significant changes with increasing $\zeta$ values. The decrease in the opening angle of the cone with increasing $\zeta$ is also evident from the changes at the shock locations of the shock cone.
    }
\label{BH2_VelocCut}
\end{figure*}

In Model-II, under strong gravitational fields, the behavior induced by the modified gravity with pressure forces included, as shown in Figs.\ref{BH2_denCut} and \ref{BH2_Norm_Avg_Acc}, is also reflected in the radial and azimuthal velocity profiles presented in Fig.\ref{BH2_VelocCut}. The left panel of Fig.\ref{BH2_VelocCut} demonstrates the azimuthal variation of the radial velocity at $r = 2.66M$ for different $\zeta$ values. As $\zeta$ increases, the inflow becomes faster toward the BH, meaning that more matter is drawn inward along the accretion line. Compared with the Schwarzschild case, this leads to a brighter photon sphere around the BH. The shock locations, which mark the boundaries of the shock cone, are observed to occur around $\phi \approx 0.5$ rad. However, the behavior of azimuthal velocity at the same radius shows that, relative to Schwarzschild, increasing $\zeta$ only slightly shifts the magnitude, with larger $\zeta$ producing marginally higher shear across the cone.

The enhanced infall velocities in Model-II represent a significant departure from the behavior observed in Model-I, where quantum corrections generally led to reduced accretion efficiency. This enhancement suggests that Model-II quantum corrections modify the effective potential in a way that deepens the gravitational well, drawing matter more efficiently toward the BH. Such effects could manifest observationally as enhanced luminosity from the inner regions of accretion disks, providing a clear signature for distinguishing Model-II from classical general relativity.

The stability of shock locations at $\phi \approx 0.5$ rad across different $\zeta$ values indicates that the large-scale geometry of the shock cone remains largely preserved in Model-II, despite the significant modifications to density and velocity profiles. This geometric stability contrasts with the progressive narrowing observed in Model-I and suggests that Model-II provides a more predictable framework for understanding quantum corrections to BH accretion.

The velocity field analysis for Model-II reveals several important features that distinguish it from both classical BH and Model-I systems. The enhanced radial velocities indicate more efficient matter transport toward the BH, while the modified azimuthal velocity patterns suggest changes in angular momentum transport that could affect disk stability and jet production. These modifications provide multiple observational channels for testing Model-II predictions through X-ray timing studies and spectral analysis of accreting systems.

The evolution of velocity patterns with $\zeta$ in Model-II follows a more predictable trend than the complex, sometimes chaotic behavior observed in Model-I. This predictability could make Model-II easier to constrain observationally, as systematic parameter estimation studies could more reliably extract the quantum correction parameter from observational data.

\subsection{Distinct Numerical Signatures of Model-I and Model-II}
\label{Compare_BH1_BH2}

The analysis of Model-I and Model-II reveals fundamental differences in how quantum corrections manifest in astrophysical observables, providing distinct pathways for distinguishing these models from classical GR and from each other \cite{isz52,isz53}. The key discriminating feature lies in the behavior of the stagnation point within the shock cone structure, which directly controls the QPO generation mechanism and observational signatures.

The stagnation point represents the location inside the shock cone where the radial velocity becomes zero. Matter located closer to the BH than this point falls inward, while matter outside of it is deflected outward. Thus, the stagnation point controls the size of the cavity formed inside the shock cone. Within this cavity, QPOs can be excited, and it regulates how energy, angular momentum, and brightness emerge from the system \cite{isz54}. The cavity size determines the characteristic timescales for matter to traverse the shock region, directly influencing the frequencies of oscillatory modes that can be sustained within the system.

\begin{figure*}[!htp]
\centering
\includegraphics[width=16.0cm,height=10.0cm]{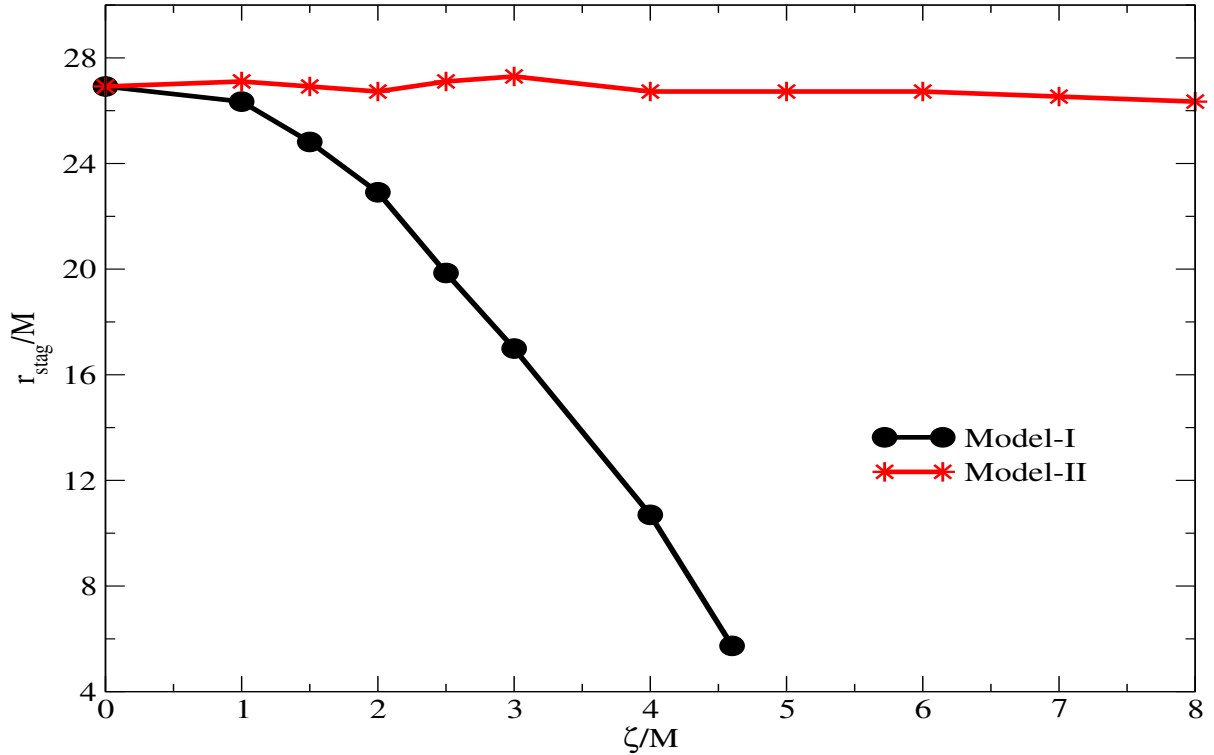}
    \caption{\footnotesize The variations of the stagnation point inside the shock cone with respect to $\zeta$ in the Model-I and Model-II models. In the Model-I case, the stagnation point rapidly approaches the BH horizon with increasing $\zeta$ values, while in the Model-II case, the stagnation point remains almost constant around $r=26.8M$.
   }
\label{BH1BH2_Stag_point}
\end{figure*}

Fig.~\ref{BH1BH2_Stag_point} shows the variation of $r_{stag}$ with respect to $\zeta$, revealing one of the most striking differences between the two QCBH models. This comparison provides perhaps the most dramatic illustration of how different types of quantum corrections can lead to completely different physical behaviors, even when both models are derived from similar underlying principles of quantum gravity.

In Model-I, the stagnation point decreases dramatically from about $27M$ to nearly $5M$ as $\zeta$ increases from $0$ to $8$, following what appears to be an exponential decay pattern. This rapid inward migration reflects the fundamental modification of both temporal and radial metric components in Model-I, which progressively alters the effective potential landscape and geodesic structure. The black circles in Fig.~\ref{BH1BH2_Stag_point} trace this steep decline, showing that by $\zeta \approx 4M$, the stagnation point has moved to within $\sim 10M$ of the horizon, representing a reduction of more than 60$\%$ from its initial position.

This significant migration greatly impacts the physics governing accretion flows. As the stagnation point moves inward, the effective size of the shock cavity decreases, reducing the volume available for matter accumulation and oscillatory mode excitation. The exponential nature of this decline suggests a threshold behavior where small increases in $\zeta$ can lead to large changes in system properties, potentially making Model-I systems highly sensitive to small variations in the quantum correction parameter.

In stark contrast, Model-II maintains a nearly constant stagnation point at $r \approx 26.8M$ throughout the entire $\zeta$ range, as shown by the red crosses in Fig.~\ref{BH1BH2_Stag_point}. This stability stems from the fact that only the radial metric component $g(r)$ is modified in Model-II while the temporal component $f(r)$ remains Schwarzschild-like. The preservation of the classical time dilation effects ensures that the overall flow dynamics and stagnation point location are largely unaffected by quantum corrections.

The stability of the stagnation point in Model-II represents a fundamentally different approach to quantum corrections, where modifications to spatial geometry do not disrupt the large-scale flow patterns that govern accretion dynamics. This stability suggests that Model-II could maintain consistent observational properties across a wide range of quantum correction parameters, potentially making it more difficult to detect but also more predictable in its behavior.

Since the stagnation point defines the cavity size, in Model-II the cavity stays large as $\zeta$ increases, which supports long advection timescales, so QPOs appear at lower characteristic frequencies and remain stable \cite{isz55}. In Model-I, however, the cavity shrinks with $\zeta$, shortening the crossing time and allowing higher-frequency oscillations to be excited. As $r_{stag}$ approaches the BH horizon, the BHL mechanism becomes inefficient, leading to less matter accreted into the BH. This removes the condition that sustains QPO generation and significantly reduces their observability. It is also worth noting that the relationship between cavity size and QPO characteristics provides a direct connection between fundamental spacetime properties and observable astrophysical phenomena. The preservation of large cavities in Model-II suggests that this model could maintain strong QPO signals across a wide parameter range, while the shrinking cavities in Model-I indicate that QPO detectability would decrease as quantum corrections become stronger.

The physical significance of this phenomenon holds considerable importance for the field of observational astronomy. In Model-I, at small values of $\zeta$, the inner disk region produces very bright and compact emission with stronger angular-momentum transfer, but as $\zeta$ increases and the stagnation point moves closer to the horizon, the density contrast and luminosity may drop. On the other hand, in Model-II, since the cavity size remains nearly unchanged, it produces steadier luminosity and angular-momentum transfer, and the QPO frequencies can remain stable. These different evolutionary paths suggest that Model-I and Model-II systems might be distinguished not only by their QPO properties but also by their overall luminosity evolution and spectral characteristics. Model-I systems might show systematic dimming as quantum corrections increase, while Model-II systems could maintain more stable emission properties while exhibiting enhanced efficiency due to improved matter focusing.

When the numerical results are compared with the theoretical ones, in the case of Model-I, as seen in \cite{isz49,isz50,isz51,QPOmetric}, increasing $\zeta$ causes the photon ring to sharpen, the observable intensity to shift inward, and if sufficient matter can accrete toward the BH, this leads to the formation of a brighter region. However, as shown in Fig.~\ref{BH1BH2_Stag_point}, once the stagnation point approaches too close to the horizon, less material is accreted, which results in a reduction of the effective emissivity and makes the brightness enhancement less dramatic than what pure ray-tracing alone would suggest.

This comparison between ray-tracing predictions and hydrodynamical results highlights an important limitation of purely geometric approaches to understanding quantum-corrected BH. While ray-tracing can accurately predict the modification of photon trajectories and lensing effects, it cannot capture the full complexity of matter dynamics and the feedback effects between modified spacetime geometry and accretion flow structure.

In contrast, the results for Model-II are consistent with the theoretical predictions given in \cite{isz49,isz50,isz51,QPOmetric}. Since the stagnation point remains nearly constant, the cavity size does not change, and the shock cone together with the emission region stay almost the same. Consequently, as also indicated by the theoretical results, the photon-ring properties remain nearly unchanged. This consistency between theory and simulation in Model-II provides confidence in our numerical methodology and suggests that Model-II represents a more predictable quantum correction scenario from an observational standpoint. The agreement between different approaches for Model-II suggests that this model provides a more robust framework for making observational predictions, as results are less sensitive to the specific details of the analysis method employed. 

\begin{figure*}[!htp]
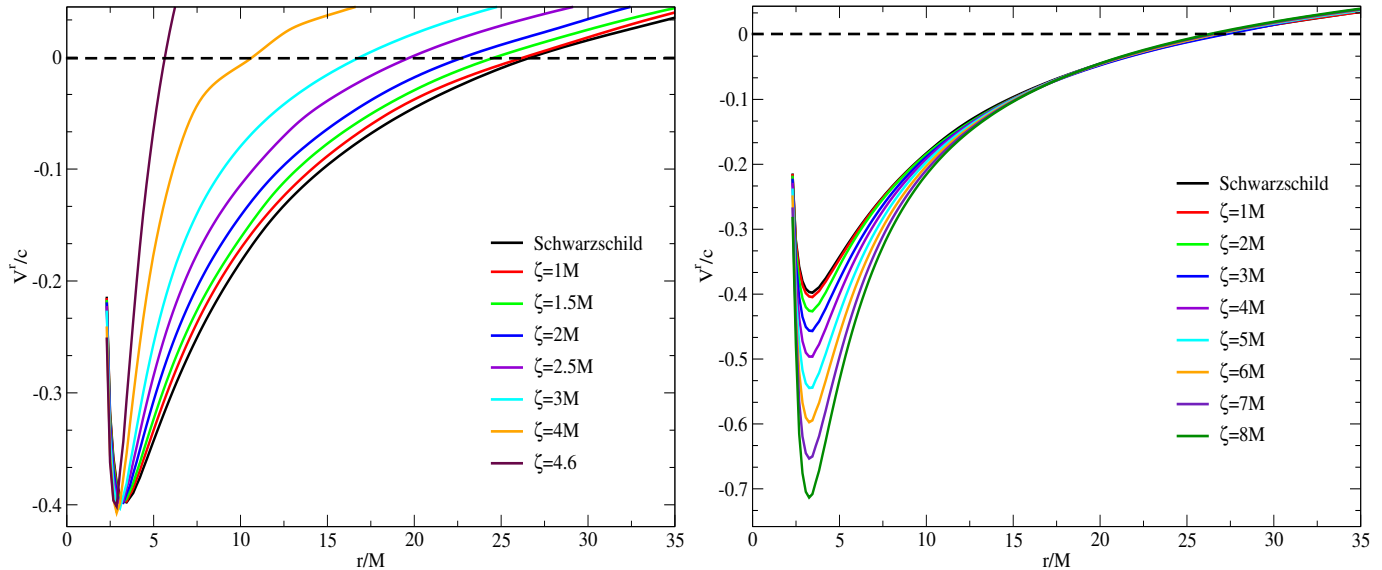

\centering
\includegraphics[width=9.0cm,height=7.5cm]{BH1radV_r_phi0.eps} 
\includegraphics[width=9.0cm,height=7.5cm]{BH2radV_r_phi0.eps} 
    \caption{\footnotesize The radial variations of the radial velocities at the fixed value $\phi=0$, corresponding to the center of the resulting shock cone, are shown for different $\zeta$ values in the Model-I and Model-II models. The plot on the left shows the variation of the radial velocity for the Model-I model, confirming that the ISCO and stagnation point progressively approach the BH horizon. However, the depth remains the same for all $\zeta$ values. In contrast to the Model-I model, in the Model-II model, the ISCO and stagnation point remain almost unchanged for each $\zeta$ value, while the depth increases with increasing $\zeta$.
   }
\label{BH1BH2_VelocCut_radial}
\end{figure*}

To compare the QCBH cases in Model-I and Model-II more quantitatively, we present the variation of the radial velocity along the radial direction at $\phi=0$ rad in Fig.~\ref{BH1BH2_VelocCut_radial}. In both panels, the dashed line indicates the location where the velocity becomes zero, corresponding to the stagnation points discussed above. This detailed analysis provides additional insight into the mechanism underlying the stagnation point behavior and its implications for matter transport and energy dissipation.

The left panel of Fig.~\ref{BH1BH2_VelocCut_radial} shows the behavior of $v^r$ as a function of $r$ for different values of $\zeta$ in the case of Model-I. As explained in Fig.~\ref{BH1BH2_Stag_point}, the intersections of the radial velocity curves with the dashed line correspond to the stagnation points. For Model-I, when the QCBH case is compared with the Schwarzschild case at different $\zeta$ values, the velocity profiles change dramatically. The Schwarzschild profile (black line) shows a stagnation point at $r \approx 27M$, while increasing $\zeta$ systematically shifts this point inward.

The systematic inward shift of velocity profiles in Model-I reflects the progressive modification of the effective potential that governs matter motion. This modification appears to create increasingly steep potential wells that draw the stagnation point closer to the BH, effectively compressing the shock cavity and altering the conditions for QPO generation. At the same time, the minimum value of $v^r$ shifts closer to the BH horizon. In particular, for $\zeta \geq 4$, the stagnation point moves very close to the horizon. The velocity profiles also show that the depth of the velocity minimum (around $r \sim 5-10M$) remains approximately constant at $v^r/c \approx -0.4$ across all $\zeta$ values, but the radial location where this minimum occurs shifts inward. Moreover, the fact that the zero-crossings of the radial velocity curve approach the horizon indicates that accretion toward the BH at large radii is significantly reduced. At very high $\zeta$, the radial inflow weakens close to the BH, signaling the breakdown of steady BHL accretion. The preservation of velocity depth while the location shifts inward suggests that the fundamental energetics of the accretion process remain similar, but the spatial distribution of these processes is compressed toward smaller radii. 

The behavior of the radial velocities for the Schwarzschild and Model-II QCBH cases at different values of $\zeta$ is shown in the right panel of Fig.~\ref{BH1BH2_VelocCut_radial}. With increasing $\zeta$, the point at which the velocity reaches its minimum becomes deeper, indicating that matter falls more rapidly toward the BH in the range $r \sim 4-6M$. The velocity minimum deepens from approximately $v^r/c \approx -0.4$ for Schwarzschild to $v^r/c \approx -0.7$ for $\zeta = 8M$, representing a 75$\%$ increase in infall speed.

The enhancement of infall velocities in Model-II while maintaining stable stagnation points represents an optimal scenario for accretion efficiency. The faster infall speeds indicate more efficient matter transport through the inner regions of the flow, while the stable large-scale structure ensures that this enhanced transport can be sustained over long periods without the instabilities that plague Model-I at large quantum corrections. However, the overall behavior of the radial velocity remains almost the same across all models. At the same time, the stagnation point appears at nearly the same location for all values of $\zeta$, consistently occurring around $r \approx 27M$. In the Model-II case, the radial velocity profiles generate a similar asymptotic structure, meaning that the cavity size formed inside the shock cone remains nearly unchanged. This stability in the large-scale flow pattern, combined with enhanced infall velocities in the inner region, creates optimal conditions for sustained QPO generation.

The combination of enhanced efficiency and structural stability in Model-II suggests that this quantum correction scenario could lead to BH systems that are both more luminous and more stable than their classical counterparts. Such systems might represent optimal targets for detecting quantum gravitational effects through long-term monitoring programs. As discussed earlier, and consistent with theoretical expectations, in Model-I the photon sphere contracts slightly while the ISCO moves outward \cite{isz51}. This tension affects how much matter can linger in circular orbits. In contrast, in Model-II, both the photon sphere and the ISCO remain almost unchanged. Therefore, as clearly shown before, in Model-I the increase in $\zeta$ produces strong deviations from GR, significantly affecting the geodesic structure and hydrodynamic infall around the BH. Moreover, in Model-I the radial velocity profile becomes shallower and the stagnation point shifts closer to the BH. This shrinks the cavity that traps QPO modes, initially leading to higher-frequency QPOs. However, as the stagnation point moves even closer to the BH, the efficiency of matter inflow through the BHL mechanism decreases significantly, the emission drops, and the brightness enhancement predicted by pure ray-tracing is suppressed. This explains the apparent discrepancy between theoretical predictions and observational expectations for Model-I in the large $\zeta$ regime.

The evolution of QPO frequencies with cavity size provides a direct test of the quantum correction models. Model-I predicts that QPO frequencies should increase as quantum corrections strengthen (due to smaller cavities) before eventually disappearing as the system becomes unstable. Model-II predicts more stable QPO frequencies that should remain observable across a wider parameter range. However, Model-II sustains a consistent cavity, so QPO frequencies remain low and stable, and the brightness is steadier, although QPOs can still be modulated by the influence of hydrodynamic forces. The details of QPO generation and excitation mechanisms are discussed extensively in Section \ref{GRHD4}. On the other hand, in the case of Model-I, both numerical simulations and theoretical calculations show that the photon ring sharpens and the intensity shifts inward with increasing $\zeta$. However, although the theoretical results indicate that this change should enhance brightness and shift variability to higher frequencies, the inward movement of the stagnation point within the cavity toward the BH leads to reduced accretion and suppressed emissivity, explaining why ray-tracing tends to overpredict brightness.

In contrast to Model-I, the results obtained for Model-II show close agreement with the theoretical predictions. This consistency arises because the temporal metric component remains unmodified, preserving the classical energy and angular momentum relationships that govern the large-scale flow structure. The quantum corrections in Model-II primarily affect the spatial geometry without fundamentally altering the causal structure or energy extraction mechanisms. These distinct behaviors provide clear observational discrimination criteria between Model-I and Model-II. Besides, Model-I should exhibit strong $\zeta$-dependent variability in both QPO frequencies and luminosity, with a tendency toward suppressed emission at large quantum corrections. Model-II should show more stable QPO characteristics with enhanced but consistent emission levels. The predictable behavior of Model-II makes it an attractive candidate for systematic observational studies, as its parameters could be constrained through relatively straightforward fitting procedures applied to X-ray timing and spectral data. The more complex behavior of Model-I might require more sophisticated analysis techniques but could provide stronger constraints once detected. Future X-ray timing missions with sufficient sensitivity could potentially distinguish between these scenarios and constrain the nature of quantum corrections in strong gravitational fields.

\subsection{QPOs Driven by Shock Cone Instabilities in Model-I and Model-II}
\label{GRHD4}

The generation of QPOs through shock cone instabilities represents one of the most promising mechanisms for observationally distinguishing QCBHs from classical BH \cite{isz52,isz53}. The temporal evolution of mass accretion rates and the resulting power spectral density (PSD) analyses provide direct windows into the modified spacetime geometry effects, revealing characteristic frequency patterns that depend sensitively on the quantum correction parameter $\zeta$. This approach combines the robust theoretical foundation of relativity with the practical accessibility of X-ray timing observations, creating a powerful tool for testing quantum gravity theories in astrophysical contexts.

The QPO generation mechanism in our simulations arises from the interaction between shock cone instabilities and the cavity structures that form within the accretion flow. These cavities act as resonant chambers that can trap and amplify various oscillatory modes, leading to the characteristic frequency patterns observed in X-ray light curves from accreting BH and neutron stars.

\subsubsection{QPO Generation in Model-I}

\begin{figure*}[!htp]
\centering
\includegraphics[width=16.0cm,height=10.0cm]{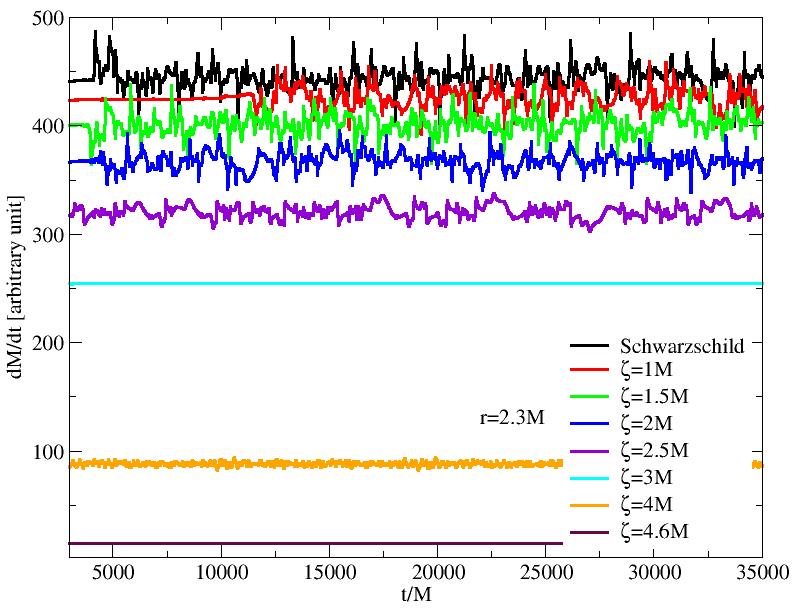}
     \caption{\footnotesize For the Model-I model, the variation of the mass accretion rate with respect to the quantum correction parameter $\zeta$, along with the Schwarzschild solution, has been shown at $r=2.3M$. It has been observed that as the value of $\zeta$ increases, there is a noticeable decrease in the amount of matter falling into the BH. Additionally, after reaching a steady state, cases exhibiting quasi-periodic instabilities have been observed to transition into a strongly stable state for $\zeta \geq 3M$, and therefore no longer exhibit QPO oscillations.}
\label{BH1_AccRate}
\end{figure*}

The mass accretion rate is an important physical mechanism for characterizing the behavior of matter around a BH. In Fig.~\ref{BH1_AccRate}, the time evolution of the mass accretion rate for different $\zeta$ values in the Model-I case is shown. It is observed that as $\zeta$ increases, the mass accretion rate steadily decreases compared to the Schwarzschild BH. For small $\zeta$ values, the instabilities are strong, while for $\zeta=3$ and larger values, the system becomes strongly stable and the accretion is suppressed.

The systematic evolution of accretion rate behavior with $\zeta$ provides crucial insight into the threshold mechanisms that govern QPO generation in quantum-corrected spacetimes. The transition from unstable to stable behavior reflects the changing balance between gravitational attraction, quantum corrections, and hydrodynamic forces, with the quantum modifications eventually overwhelming the instability mechanisms that drive QPO formation.

\begin{figure}[ht!]
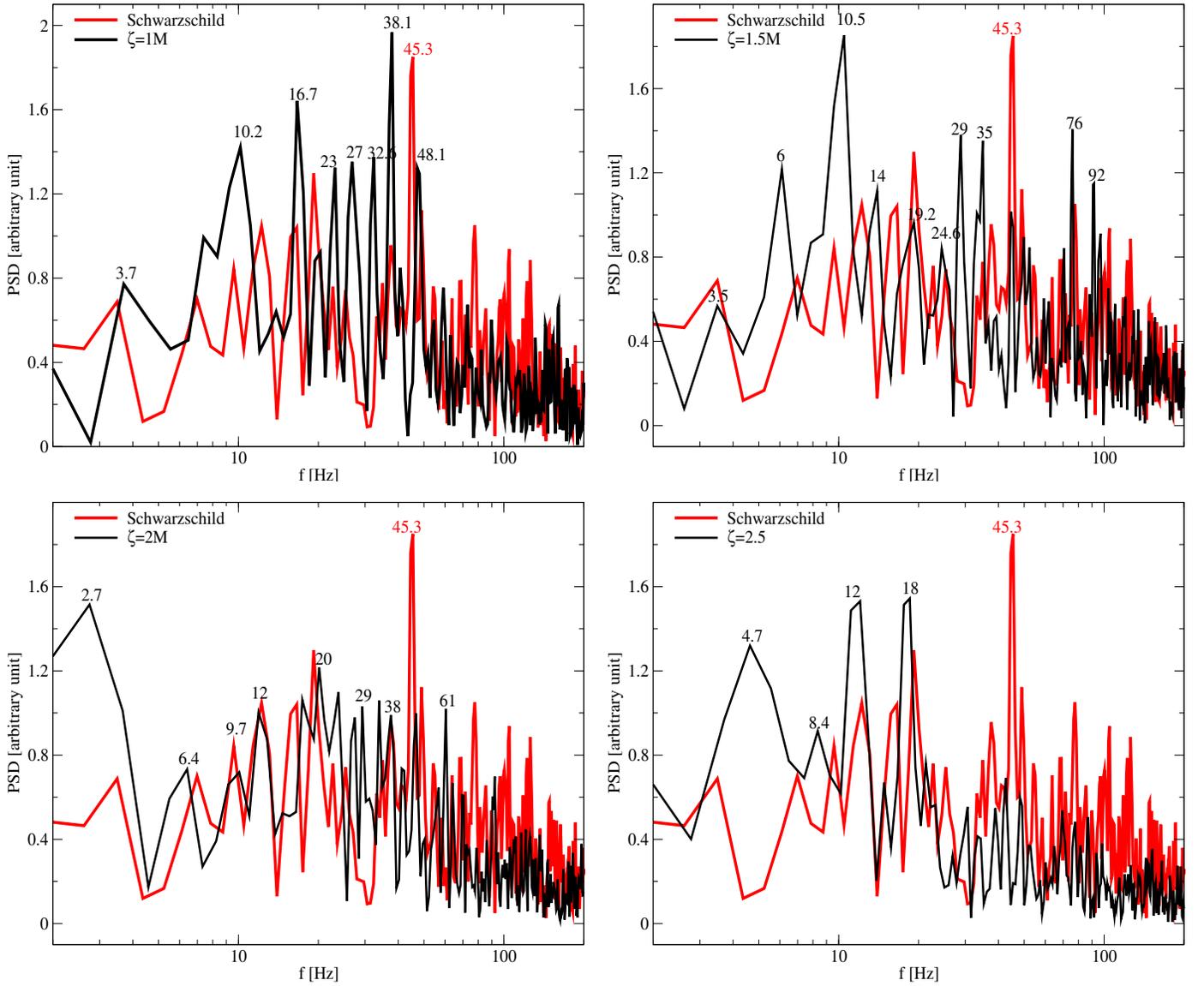

\centering
\includegraphics[width=9.0cm,height=7.5cm]{PSDBH1zeta1.eps}\;\; 
\includegraphics[width=9.0cm,height=7.5cm]{PSDBH1zeta15.eps} \\
\vspace{0.2cm}
\includegraphics[width=9.0cm,height=7.5cm]{PSDBH1zeta2.eps}\;\; 
\includegraphics[width=9.0cm,height=7.5cm]{PSDBH1zeta25.eps} 
    \caption{\footnotesize For the Model-I model at different $\zeta$ values, PSD analyses are presented in comparison with the Schwarzschild case for $M=10M_\odot$, but only for the cases where quasi-periodic oscillations are observed ($\zeta<3$). As expected, due to the change in the physical structure of the shock cone formed around the BH with varying $\zeta$ values, the positions of the resulting QPO peaks have shifted. It has been observed that with increasing $\zeta$, low-frequency peaks are excited.
   }
\label{BH1_QPOs}
\end{figure}

The critical transition occurs around $\zeta = 2-2.5M$, where the oscillations become increasingly damped. For $\zeta \geq 3M$, the system enters a stable regime with minimal fluctuations, corresponding to the suppression of shock cone instabilities. This behavior directly correlates with the stagnation point migration shown in Fig.~\ref{BH1BH2_Stag_point}, confirming that cavity shrinkage leads to QPO suppression \cite{isz56}. The identification of this critical threshold provides an important constraint on the observability of quantum corrections in astrophysical systems. The fact that QPOs disappear for $\zeta \geq 3M$ suggests that strongly quantum-corrected BH might appear observationally as stable, non-variable sources, potentially providing a distinctive signature for identifying such systems in astronomical surveys. As a result, the instabilities clearly visible in the mass accretion rate represent signals indicating the formation of QPOs. As will be discussed in Fig.~\ref{BH1_QPOs}, QPOs arise for $\zeta<3$, whereas for $\zeta \geq 3M$, due to the strongly stable behavior over time, no QPOs are formed. As seen in Fig.~\ref{BH1_AccRate}, once the plasmatic structure around the BH and thus the shock cone reach a steady-stable configuration, it is observed to generate instabilities. Consequently, the fundamental radial and azimuthal modes theoretically shown in Fig.~\ref{f1} become trapped inside the shock cone and give rise to QPO frequencies. 

As discussed in the previous works \cite{isz52,isz53}, these fundamental modes undergo nonlinear coupling, producing a large number of peaks in the PSD analysis. The connection between theoretical epicyclic frequencies and numerically observed QPOs provides a crucial validation of our approach. The ability to predict QPO characteristics from first principles, starting with the modified spacetime geometry and proceeding through hydrodynamical simulations to observable frequency spectra, demonstrates the power of combining analytical and numerical methods for testing quantum gravity theories.

This situation is illustrated for Model-I in Fig.~\ref{BH1_QPOs} at different values of $\zeta$. In Fig.~\ref{BH1_QPOs}, the first peaks that appear in each case represent the fundamental modes, while the subsequent peaks arise either from nonlinear couplings among the fundamental modes themselves or from further couplings between newly formed peaks and the fundamental modes. In each $\zeta$ case shown in Fig.~\ref{BH1_QPOs}, the PSD analysis is plotted in comparison with that obtained for the Schwarzschild BH.

The PSD evolution in Fig.~\ref{BH1_QPOs} reveals several key features. For $\zeta = 1M$ (upper left panel), multiple peaks emerge across both LFQPO and HFQPO bands, with prominent features at 16.7 Hz, 32.6 Hz, and 48.1 Hz. The $3:2$ ratio between 48.1 and 32.6 Hz, and the $2:1$ ratio between 32.6 and 16.7 Hz, suggest strong nonlinear mode coupling \cite{isz57}. As $\zeta$ increases to 1.5M (upper right panel), the spectral content shifts toward lower frequencies, with new prominent peaks at 14 Hz, 19.2 Hz, and 29 Hz in the LFQPO band, while maintaining HFQPO features at 45.3 Hz and 92 Hz. The systematic frequency migration with $\zeta$ reflects the changing cavity size and resonance conditions as the stagnation point moves inward. For $\zeta = 2M$ (lower left panel), the spectrum becomes dominated by LFQPO features, with multiple peaks between 6-20 Hz showing various harmonic relationships. The emergence of lower-frequency modes indicates that the shrinking cavity preferentially excites longer-period oscillations. Finally, at $\zeta = 2.5M$ (lower right panel), only a few strong LFQPO peaks remain, signaling the approach to the stable regime.

The evolution from HFQPO-dominated to LFQPO-dominated spectra provides a clear prediction for how quantum corrections should manifest in observational data. This evolution could potentially be used to constrain the quantum parameter through systematic studies of QPO properties in large samples of BH systems.

As can be seen, the excitation frequencies and their amplitudes differ entirely from those in the Schwarzschild case. The reason, as discussed above, lies in the changes in the shock-cone structure: with increasing $\zeta$, not only does the radial cavity size shrink, but the cone opening angle also decreases. These structural modifications strongly affect both the fundamental QPO frequencies and their nonlinear couplings. As shown in Fig.~\ref{BH1BH2_Stag_point}, the stagnation point gradually approaches the BH horizon, and as a result the BHL accretion mechanism weakens. Consequently, for $\zeta \geq 3M$, as also seen in Fig.~\ref{BH1_AccRate}, the shock cone becomes more strongly stabilized, which in turn leads to the disappearance of QPOs.

In every case of Fig.~\ref{BH1_QPOs}, both LFQPOs (1–30 Hz) and HFQPOs ($>30$ Hz) appear. However, it is clearly seen that as $\zeta$ increases, the QPO frequencies shift while the LFQPOs become more strongly excited. A detailed inspection of the graphs and corresponding QPO frequencies shows the following commensurate ratios: at $\zeta=1M$, $48.1:32.6 \approx 3:2$ and $32.6:16.7 \approx 2:1$; at $\zeta=1.5M$, $29:19.2 \approx 3:2$, $92:45.3 \approx 2:1$, and $29:14 \approx 2:1$; at $\zeta=2M$, $9.7:6.4 \approx 3:2$, $20:9.7 \approx 2:1$, $20:12 \approx 5:3$, and $45.3:29 \approx 3:2$; and at $\zeta=2.5M$, $18:12 \approx 3:2$ and $18:8.4 \approx 2:1$. These ratios occur with error margins in the range of 0–7$\%$. The preservation of simple integer ratios across different $\zeta$ values suggests that the underlying resonance mechanisms remain fundamentally similar to those operating in classical BH, even as the specific frequencies are modified by quantum corrections. This preservation provides confidence that our quantum-corrected models maintain physical consistency while introducing new observable signatures.

The numerical QPO frequencies presented in Fig.~\ref{BH1_QPOs} are consistent with the $\zeta$-dependent variations of the radial and azimuthal frequencies shown in Fig.~\ref{f1}. As $\zeta$ increases, $\nu_r$ changes and the characteristic radii shift slightly. This leads to $\zeta$-dependent modifications in the resonance conditions (e.g., $3:2$), either by changing the radial location or the corresponding frequency. At the same time, larger values $\zeta$ significantly reshape $\Omega_p$ near the ISCO, which is again consistent with the shift or the emergence of additional commensurate peaks in the PSD. When comparing the QPO frequencies obtained in Fig.~\ref{BH1_QPOs} for $M=10M_\odot$ with observations from XRBs, it is found that the QPOs emerging at different values of $\zeta$ are consistent with the observational data \cite{isz40,isz41}. For example, in the case of $\zeta=1M$, the peaks occurring in the 1–30 Hz range fall within the LFQPO band, while the other peaks correspond to HFQPOs. Both these frequencies and the resulting $3:2$ resonance are consistent with the resonance observed in the source GRS 1915+105 (67:41 Hz) \cite{isz47,isz48,isz40,isz43}.

The agreement between our numerical predictions and observational data from well-studied X-ray binary systems provides strong validation of our theoretical framework. This consistency suggests that the quantum correction models are not only mathematically self-consistent but also capable of reproducing the essential physics of real astrophysical systems. Thus, the simulations demonstrate both low-frequency ($\sim 3-30$ Hz) and near-HF ($\sim 45$ Hz) peaks, suggesting a Schwarzschild-like baseline with additional low-frequency excitations. For $\zeta=1.5M$ the numerically obtained $92:45.3 \approx 2:1$ ratio is similar to the HFQPOs reported in GRS 1915+105 \cite{isz47,isz48,isz40}, while the $29:19.2 \approx 3:2$ ratio agrees with type-C LFQPOs observed in XRBs \cite{isz41}. At $\zeta=2M$, the numerically predicted 6-20 Hz range matches well with the type-B/C LFQPOs observed in XTE J1550–564 \cite{isz42,isz46,isz44} and GX 339–4 \cite{isz41,isz45}, where $3:2$ and $2:1$ ratios have also been reported. Finally, in the case of $\zeta=2.5M$, the resonant states found in the simulations are again consistent with those observed in the sources mentioned above.

In summary, for Model-I, $\zeta \simeq 1-1.5M$ produces multiple peaks that result in a broad-band spectrum, resembling the mixed LFQPO and harmonic structures typical of the hard/intermediate states in transient XRBs. At intermediate values such as $\zeta=2M$, multiple commensurate ratios emerge, strongly reminiscent of type-C LFQPOs \cite{isz41} with harmonics. At high values of $\zeta$, a very clear $3:2$ ratio is obtained, consistent with the observed HFQPOs. Finally, for $\zeta \geq 3M$, the disappearance of QPOs is qualitatively consistent with stable, disk-dominated soft states in which LFQPOs are no longer present. This evolution of QPO properties with quantum corrections provides a framework for interpreting the diversity of QPO phenomena observed in different BH systems. The model suggests that variations in QPO characteristics might reflect not only differences in mass, spin, or accretion rate, but also the strength of quantum gravitational effects operating in different systems.

\subsubsection{QPO Generation in Model-II}

\begin{figure*}[!htp]
\centering
\includegraphics[width=16.0cm,height=10.0cm]{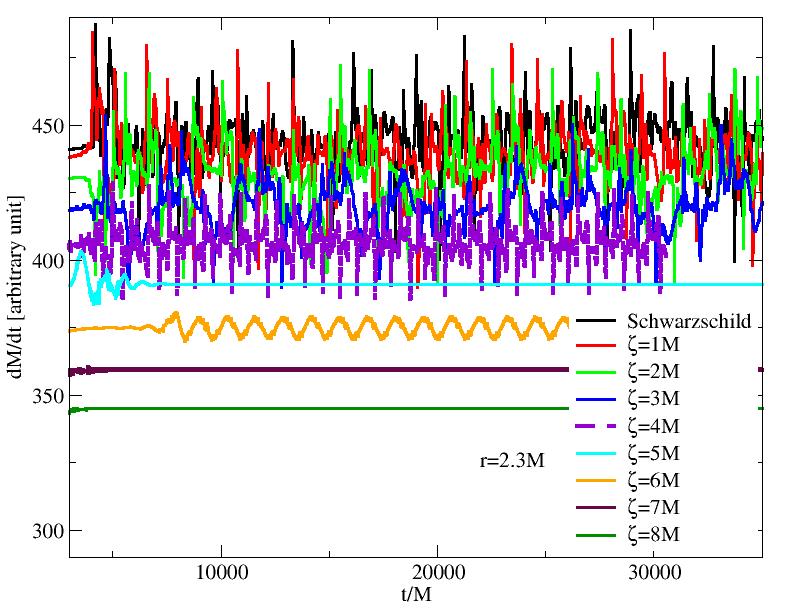}
    \caption{\footnotesize The figure is the same as Fig.~\ref{BH1_AccRate}, but this time it shows the variation of the mass accretion rate for the Model-II case. Unlike Model-I, the decrease in the mass accretion rate with increasing $\zeta$ values is not as pronounced as in Model-I. However, the amount of matter falling into the BH still decreases with increasing $\zeta$. In the Model-II case, it is observed that starting from $\zeta \geq 5$, the matter around the BH transitions into a strongly stable state.}
\label{BH2_AccRate}
\end{figure*}

As in Fig.~\ref{BH1_AccRate} for Model-I, this time the mass accretion rate for the Model-II case is presented in Fig.~\ref{BH2_AccRate}, showing its variation for different values of $\zeta$. Compared with Model-I, the change in the mass accretion rate with respect to $\zeta$ is milder. The instabilities are weaker and remain very similar to the Schwarzschild case. Only for $\zeta \geq 5M$, the matter around the BH exhibits a more stable behavior, which leads to the suppression of oscillations and the disappearance of QPOs.

The more gradual evolution of accretion properties in Model-II reflects the fundamentally different nature of quantum corrections in this scenario. The preservation of classical temporal relationships ensures that the large-scale dynamics remain stable even as spatial geometry undergoes significant modifications, leading to a more predictable and controllable evolution of system properties.

In the Model-II case, although, as seen in Fig.~\ref{BH1BH2_Stag_point}, the stagnation point remains nearly constant, the reduction in the shock cone opening angle with increasing $\zeta$ (as shown in Figs.~\ref{BH2_denCut} and \ref{BH2_VelocCut}) causes the cone to enter a strongly stable configuration. The time evolution in Fig.~\ref{BH2_AccRate} shows much more gradual changes compared to Model-I, with oscillatory behavior persisting to higher $\zeta$ values. The persistence of oscillatory behavior to higher $\zeta$ values in Model-II provides a longer observational window for detecting quantum corrections.  

\begin{figure}[ht!]
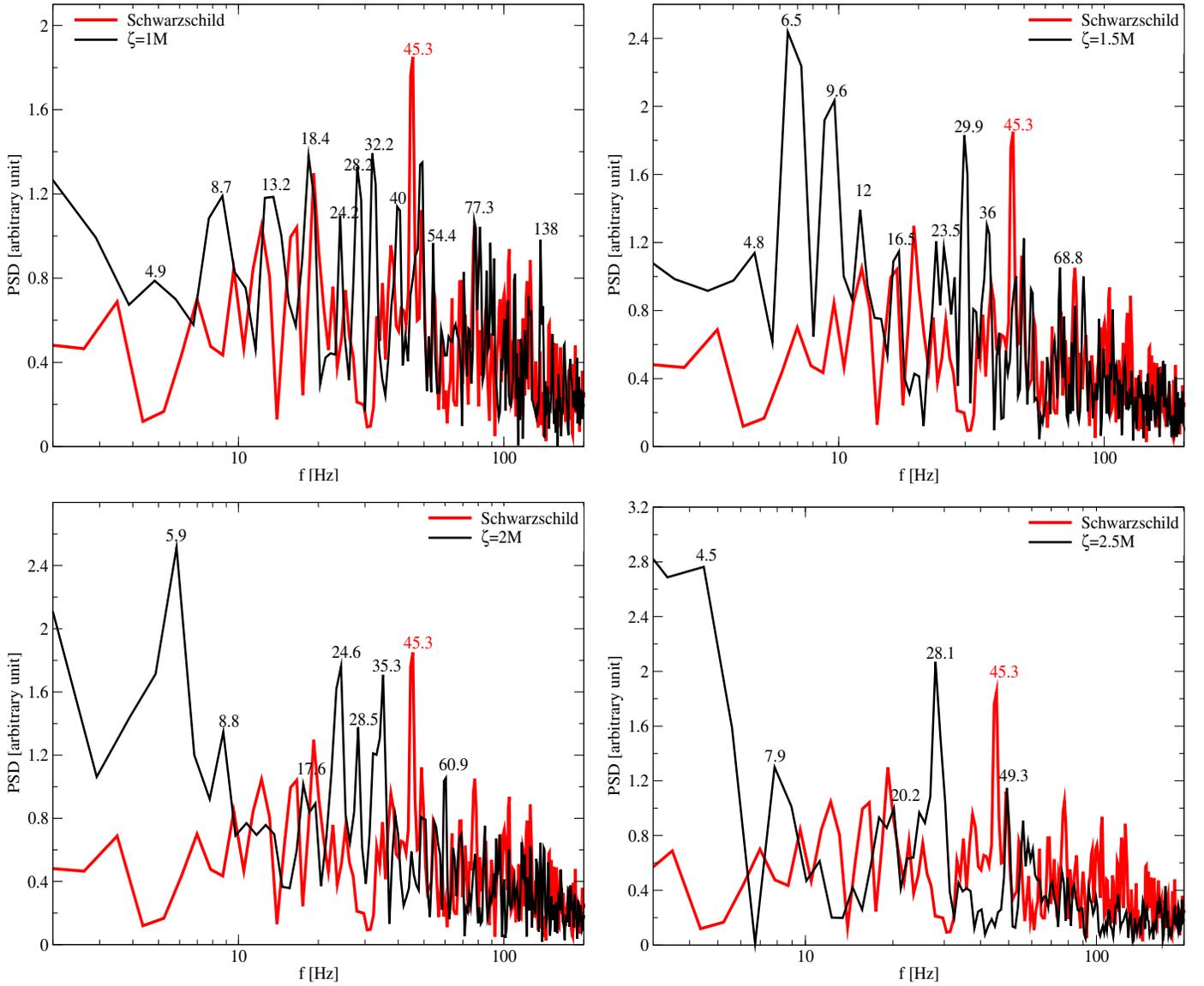

\centering
\includegraphics[width=9.0cm,height=7.5cm]{PSDBH2zeta1.eps}\;\; 
\includegraphics[width=9.0cm,height=7.5cm]{PSDBH2zeta15.eps} \\
\vspace{0.2cm}
\includegraphics[width=9.0cm,height=7.5cm]{PSDBH2zeta2.eps}\;\; 
\includegraphics[width=9.0cm,height=7.5cm]{PSDBH2zeta25.eps} 
    \caption{\footnotesize The variation of the PSD analysis with respect to different $\zeta$ values has been compared with the Schwarzschild case, assuming the BH has a mass of $M=10M_\odot$, for the Model-II model. For each $\zeta$ value, it has been observed that the resulting QPO frequencies change due to the dynamic changes in the physical mechanism, and they are also excited accordingly. In other words, it has been observed that low-frequency QPOs are excited as $\zeta$ increases.
   }
\label{BH2_QPOs_1}
\end{figure}

\begin{figure}[ht!]
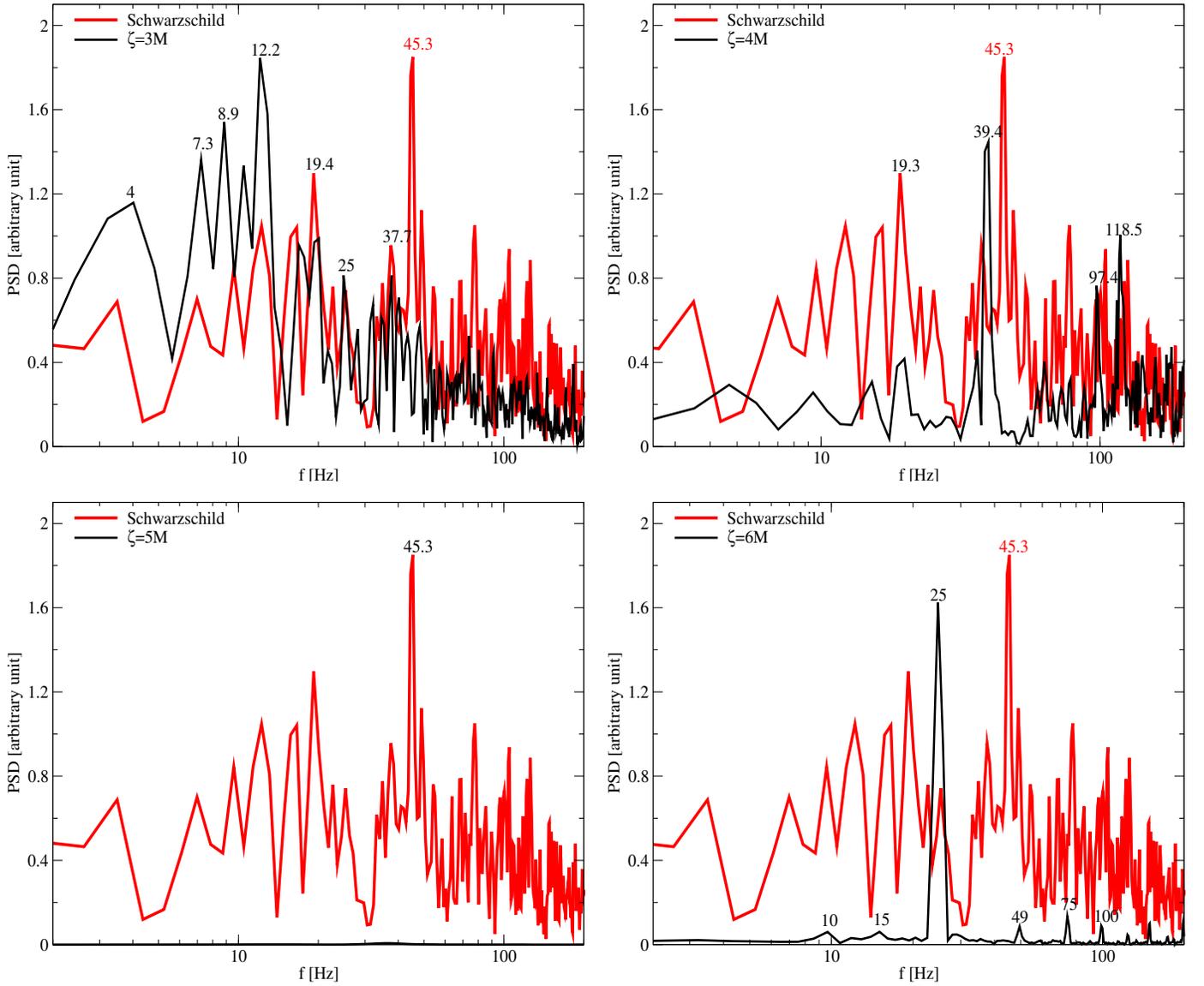

\centering
\includegraphics[width=9.0cm,height=7.5cm]{PSDBH2zeta3.eps}\;\;
\includegraphics[width=9.0cm,height=7.5cm]{PSDBH2zeta4.eps} \\
\vspace{0.2cm}
\includegraphics[width=9.0cm,height=7.5cm]{PSDBH2zeta5.eps}\;\; 
\includegraphics[width=9.0cm,height=7.5cm]{PSDBH2zeta6.eps} 
    \caption{\footnotesize For the model in Fig.~\ref{BH2_QPOs_1}, different $\zeta$ parameters have been considered, and as $\zeta$ increases, the resulting QPO frequencies are revealed. In the Model-II model case, for $\zeta<4$, low-frequency QPOs are more strongly excited, while at $\zeta=4$, high-frequency QPOs become more dominant. It has been observed that for $\zeta>4$, no QPOs are formed.
   }
\label{BH2_QPOs_2}
\end{figure}

The structure of the stagnation point, and consequently the resulting cavity, behaves completely differently in Model-II compared to Model-I, which strongly affects the trapped $\nu_r$ and $\nu_{\phi}$ frequencies inside the shock cone and their nonlinear couplings. In Figs.~\ref{BH2_QPOs_1} and \ref{BH2_QPOs_2}, PSD analyses for different values of $\zeta$ in the Model-II case are presented in comparison with the Schwarzschild case.

In Fig.~\ref{BH2_QPOs_1}, the PSD analysis for four different $\zeta$ values is shown against the Schwarzschild baseline. For $\zeta<3M$, the behavior of the peaks resembles the Schwarzschild baseline; however, as in Model-I, low-frequency QPOs are more strongly excited. As $\zeta$ increases, the PSD broadens, the frequency positions of the peaks shift, and LFQPOs become more dominant. This enhances the likelihood of their detectability with X-ray telescopes.

The evolution pattern in Model-II differs markedly from Model-I. For $\zeta = 1M$ (upper left), the spectrum closely resembles Schwarzschild with additional low-frequency enhancement around 8-18 Hz. The preservation of the large cavity size allows both fundamental and harmonic modes to coexist effectively. At $\zeta = 1.5M$ (upper right), new spectral features emerge while maintaining the overall structure, indicating stable mode excitation. For $\zeta = 2M$ and $\zeta = 2.5M$ (lower panels), the spectra show increased complexity with multiple peaks in the LFQPO band, but without the dramatic frequency migration observed in Model-I. This stability reflects the constant stagnation point behavior shown in Fig.~\ref{BH1BH2_Stag_point}.

The preservation of spectral structure in Model-II suggests that this quantum correction scenario could maintain observable QPO signatures over extended periods, making it an attractive target for long-term monitoring campaigns designed to detect quantum gravitational effects. The stability of the frequency patterns also suggests that Model-II systems could serve as standard candles for testing quantum gravity theories.

In Fig.~\ref{BH2_QPOs_2}, the PSD analysis for $\zeta=3M$ and larger values is given compared to Schwarzschild. When Figs.~\ref{BH2_QPOs_1} and \ref{BH2_QPOs_2} are analyzed together, it is seen that for $\zeta<4M$, LFQPOs dominate and are strongly excited. At $\zeta=4M$, HFQPOs become dominant. For $\zeta>4M$, as the opening angle of the cone decreases and the structure becomes more solid, strong stability emerges and the formation of QPO completely disappears. However, at $\zeta=6M$, as also seen in Fig.~\ref{BH2_AccRate}, although the system is globally stable, local instabilities are triggered due to the strong compression of the cone structure (see Figs.~\ref{BH2_denCut} and \ref{BH2_VelocCut}). This can excite a residual mode that manifests itself as a peak.

In Fig.~\ref{BH2_QPOs_1}, a careful inspection shows that for almost every $\zeta$ value, the ratios $\sim19:\sim12 \approx 3:2$ and $\sim25:\sim 12\approx 2:1$ appear. These correspond to frequencies that fall within the LFQPO observational band (1–30 Hz). Such ratios are consistent with observational results reported for the sources XTE J1550–564 \cite{isz42} and GX 339–4 \cite{isz45} (type-B/C LFQPOs). Similarly, for intermediate HFQPOs, the numerical results also exhibit the $3:2$ ratio from $\sim37:\sim25 \approx3:2$.

In Fig.~\ref{BH2_QPOs_2}, for $\zeta=3M$, the LFQPO band shows the ratios $19.4:12.2 \approx 3:2$ and $12.2:7.3 \approx 5:3$, while at $\zeta=4M$ the HFQPO band exhibits $118.5:\sim59 \approx 2:1$. These ratios are in agreement with GRS 1915+105 observations (67:41 $\approx 3:2$, and 113:56 $\approx 2:1$) \cite{isz47}. For other $\zeta$ values in Fig.~\ref{BH2_QPOs_2}, such harmonic pairs are not observed. The preservation of characteristic frequency ratios across different quantum correction strengths provides a robust observational signature that could be used to identify and constrain Model-II systems. The consistency of these ratios with observations from multiple X-ray binary systems suggests that Model-II captures essential aspects of the QPO generation mechanism while introducing subtle but detectable modifications.

Overall, the comparison between Model-I and Model-II QPO characteristics reveals fundamental differences in how quantum corrections affect oscillatory behavior. Model-I shows rapid QPO suppression due to cavity shrinkage, while Model-II maintains stable oscillations over a wider $\zeta$ range due to preserved cavity size. These distinct signatures provide clear observational discrimination criteria for distinguishing between different quantum correction scenarios in astrophysical BH.


 \section{Comparison Between Numerical Results and Analytical Predictions and Observations}
\label{isec5}

The validation of numerical simulations against theoretical predictions and observational constraints represents a crucial step in establishing the credibility of QCBH models and their potential for distinguishing quantum gravity effects from classical GR \cite{sec5is01,sec5is02}. This section systematically compares our hydrodynamical simulation results with existing analytical studies, EHT observational constraints, and known QPO observations from XRBs, providing a comprehensive assessment of the consistency and predictive power of the quantum correction framework.

\subsection{Theoretical Validation and EHT Constraints}

\begin{figure*}[!htp]
\centering
\includegraphics[width=16.0cm,height=10.0cm]{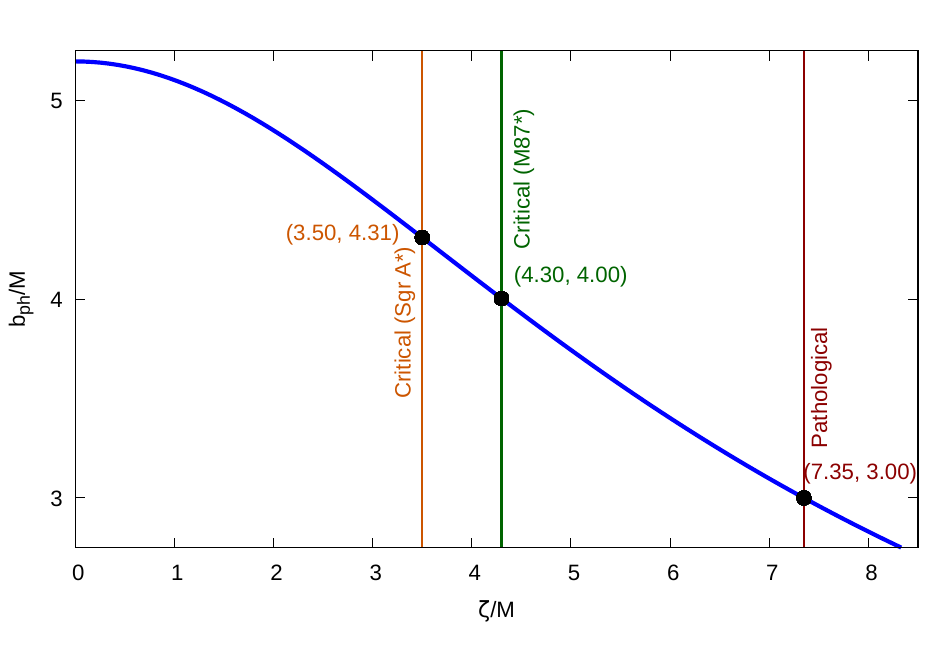}
    \caption{\footnotesize This is a graphical summary of the maximum limit values obtained for the $\zeta$ parameter, based on the comparison of the critical impact parameter of photon sphere derived analytically in the Model-I case in \cite{isz51} with the critical impact parameter observed by the EHT.
   }
\label{Function} 
\end{figure*}

Using Model-I, the effects of the parameter $\zeta$ on the BH shadow, the photon ring, and the optical appearance were investigated in \cite{isz51}. The analytical results obtained there were compared with those for a Schwarzschild BH, and $\zeta$ was also constrained using the EHT observations of $M87^*$ and $Sgr A^*$. The findings of \cite{isz51} are summarized graphically in Fig.~\ref{Function}. The paper itself emphasizes that $\zeta$ cannot be chosen arbitrarily and uses EHT shadow sizes to bound $\zeta$ for Model-I, finding $\zeta \leq 4.7M$ ($M87^*$) and $\zeta \leq 3.52M$ ($Sgr A^*$).

Consistent with this, our Fig.~\ref{BH1_Norm_Avg_Acc} shows that for $\zeta \geq 4M$ the mass accretion rate becomes severely suppressed, so the matter fails to form a plasma structure around the BH and, consequently, no shock cone develops. Therefore, the value $\zeta$ that we numerically constrain in Fig.~\ref{BH1_Norm_Avg_Acc} is in near agreement with the observation limits for $M87^*$ and $Sgr A^*$ reported in \cite{isz51}.

The remarkable convergence between our hydrodynamical constraints and the EHT-based limits provides strong validation of both approaches. Figure~\ref{Function} demonstrates that the critical impact parameter $b_{ph}/M$ decreases monotonically with $\zeta$, reaching the pathological regime around $\zeta \approx 7.35M$ where the photon sphere behavior becomes unphysical. The EHT-derived constraints place $M87^*$ at $(4.30, 4.00)$ and $Sgr A^*$ at $(3.50, 4.31)$ in the $(\zeta/M, b_{ph}/M)$ parameter space, both well within the physically allowed region.

Our numerical simulations independently arrive at similar constraints through a completely different physical mechanism: the breakdown of steady BHL accretion. The fact that both the geometric (photon sphere) and hydrodynamic (shock cone) approaches yield consistent upper bounds on $\zeta$ strengthens confidence in the physical reality of these limits \cite{isz50,isz51}.

The fact that the brightness found in \cite{isz51} increases with $\zeta$ does not contradict our suppression of the mass accretion rate, because the intensity model in \cite{isz51} scales with redshift factors (proportional to $f(r)^2$, as given in Eq.~\ref{function-1}) and with an assumed emissivity, while keeping the mass supply implicit. They do not tie image brightness to a self-consistent mass accretion rate. This distinction highlights the complementary nature of ray-tracing and hydrodynamical approaches: the former focuses on photon trajectories and gravitational lensing effects, while the latter captures the full dynamics of matter flow and shock formation.

The behavior of the rest-mass density in Fig.~\ref{BH1_denCut}, when compared with the analytic results in Fig.~\ref{Function} and \cite{isz50,isz49,isz51,QPOmetric}, shows that the consistency is striking in the qualitative sense. As seen in Figs.~\ref{fig:model1_embeddings} and \ref{Function}, the critical impact parameter $b_{ph}$ decreases with increasing $\zeta$, indicating that the photon sphere approaches the BH horizon. At the same time, as shown in \cite{isz50,isz49,isz51,QPOmetric}, the analytic calculations demonstrate that the BH shadow itself changes with $\zeta$. Alongside these changes, the width and brightness of the photon sphere are also reduced.

These behaviors are consistent with the numerical results obtained in Fig.~\ref{BH1_denCut}. In both cases (analytic and numerical studies), as $\zeta$ increases, the near-horizon structures-shock cones in the simulation and photon rings in the theoretical ray-tracing study-become weaker, narrower, and less pronounced. Moreover, both numerical and theoretical analyses reveal that with increasing $\zeta$, the brightness of the photon ring is suppressed in ray-tracing calculations, while in hydrodynamical simulations the density of the shock cone is also suppressed. Thus, the qualitative similarity lies in the weakening of horizon-proximal structures with increasing $\zeta$.

The radial and azimuthal velocity profiles at $r=2.66M$ shown in Fig.~\ref{BH1_VelocCut} confirm the trends found in \cite{isz51} that are summarized in Fig.~\ref{Function}. As $\zeta$ increases, the critical impact parameter $\beta_c$ decreases, the emission region of the photon ring becomes narrower and shifts toward the BH horizon, and the bright photon ring becomes thinner and brighter. In other words, the numerical velocity plots presented here support not only the analytical results of \cite{isz51}, but are also consistent with the $\zeta$-dependent evolution of the photon sphere around the BH shown in \cite{isz50,isz49,isz51,QPOmetric}.

\subsection{QPO Predictions for Supermassive BH}

By adapting the PSD analysis presented in Fig.~\ref{BH1_QPOs} for Model-I to the massive BH sources $Sgr A^*$ and $M87^*$, we demonstrate the consistency between the analytically constrained behavior of the $\zeta$ parameter for $Sgr A^*$ and $M87^*$ (as discussed in \cite{isz51} and graphically summarized in Fig.~\ref{Function}) and the numerically calculated QPOs shown in Fig.~\ref{BH1_QPOs}. In this way, by discussing the relationship between the possible maximum value of $\zeta$ for these massive BH sources and the numerical QPOs, we reveal the potential QPOs that could be observed from these sources.

The numerically calculated QPOs in Fig.~\ref{BH1_QPOs} for $M=10M_{\odot}$ agree well with the observations. Therefore, the shock cone and cavity formed around BH in XRBs can be proposed as an important physical mechanism for the QPOs observed in such sources \cite{isz52,isz53}. The same mechanism can also be suggested for supermassive BH like $Sgr A^*$ and $M87^*$, which host massive BH at their centers. In this case, using our numerical results, we can estimate the possible observable QPO frequencies of these sources, whose shadows have been imaged by the EHT.

For this purpose, the QPO frequencies calculated in Fig.~\ref{BH1_QPOs} are rescaled using the given formula to predict the possible QPOs around these supermassive BH,
\begin{equation}
f(M) = f_{ref}\frac{10M_{\odot}}{M}.
\end{equation}

\begin{table}
\footnotesize
\caption{\footnotesize Some of the QPOs numerically calculated in Fig.~\ref{BH1_QPOs} for $M=10M_{\odot}$ have been rescaled and computed for the supermassive BH with masses $M_{Sgr A^*} = 4.15 \times 10^6M_{\odot}$, $M_{M87^*} = 6.5 \times 10^9M_{\odot}$.}
\label{QPO_table}
\begin{center}
\vspace*{-0.2cm}
 \begin{tabular}{|c|c|c|c|c|c|c|}
   \hline
   \hline
   $\zeta$ & \textit{ratio} & $f_1 (Hz) \; for \; 10M_{\odot}$ & $f_2 (Hz) \; for \; 10M_{\odot}$ & \textit{Source} &  $f_1 (Hz)$ & $f_2 (Hz)$ \\
   \hline
    &  &  &  &  \textit{$Sgr A^*$} & $78.6 \;\mu Hz$ & $115.9 \;\mu Hz$ \\  
   $1$ & $3:2$ & $32.6$ & $48.1$ &   &   &   \\
     &  &  &  & \textit{$M87^*$} & $50.2 \;nHz$  & $74 \;nHz$  \\  
     \hline
    &  &  &  &  \textit{$Sgr A^*$} & $109 \;\mu Hz$ & $222 \;\mu Hz$ \\  
   $1$ & $2:1$ & $45.3$ & $92$ &   &   &   \\
     &  &  &  & \textit{$M87^*$} & $69.6 \;nHz$  & $141 \;nHz$  \\  
     \hline
   &  &  &  &  \textit{$Sgr A^*$} & $15.4 \;\mu Hz$ & $23.3 \;\mu Hz$ \\  
   $2$ & $3:2$ & $6.4$ & $9.7$ &   &   &   \\
     &  &  &  & \textit{$M87^*$} & $9.85 \;nHz$  & $14.9 \;nHz$  \\    
      \hline
   &  &  &  &  \textit{$Sgr A^*$} & $28.9 \;\mu Hz$ & $43.3 \;\mu Hz$ \\  
   $2.5$ & $3:2$ & $12$ & $18$ &   &   &   \\
     &  &  &  & \textit{$M87^*$} & $18.5 \;nHz$  & $27.7 \;nHz$  \\         
   \hline
   \hline
 \end{tabular}
\end{center}
\vspace*{-0.5cm}
\end{table}

For some commensurate ratios that appear in Fig.~\ref{BH1_QPOs} and are discussed in detail, the possible QPO frequencies of $Sgr A^*$ and $M87^*$ have been calculated and presented in Table~\ref{QPO_table}. As can be seen in Table~\ref{QPO_table}, although the absolute frequencies change due to the mass of the source, the resulting ratios remain the same.

The predictions in Table~\ref{QPO_table} reveal several important implications for future observational campaigns. For $Sgr A^*$, the predicted frequencies lie in the microhertz range, which is accessible to long-term X-ray monitoring campaigns with current and future X-ray missions \cite{isz36,isz37}. The $3:2$ resonance frequencies at $\zeta = 1$ predict oscillations at 78.6 and 115.9 $\mu$Hz, corresponding to periods of approximately 3.5 and 2.4 hours respectively. Such timescales are well-suited for extended observations spanning multiple orbital periods.

For $M87^*$, the predicted frequencies fall in the nanohertz range, which is also numerically confirmed by \cite{Orh6} using the parameters of the hairy Kerr BH, and would require observational campaigns spanning months to years to detect the periodic signatures \cite{isz38}. While challenging, such ultra-low-frequency variability has been observed in other supermassive BH systems and could potentially be detected through careful analysis of long-term EHT datasets or coordinated multi-wavelength monitoring \cite{isz39}.

The preservation of frequency ratios across different mass scales provides a robust test of the quantum correction hypothesis. The $3:2$ and $2:1$ ratios predicted for stellar-mass BH should appear at proportionally scaled frequencies in supermassive systems, offering a direct way to validate the quantum correction framework using observational data.


\section{\large Conclusions}\label{isec6}

We studied two simple quantum-corrected, spherically symmetric QCBH models and derived analytic constraints and expansions for the periastron-precession frequency $\Omega_p=\Omega_\phi-\Omega_r$. The comparative analysis of Model-I and Model-II reveals fundamental differences in how quantum corrections manifest in astrophysical observables, providing distinct pathways for distinguishing these models from classical GR and from each other. Our investigation spans both analytical derivations of epicyclic frequencies and comprehensive numerical simulations of BHL accretion, establishing a robust theoretical and computational framework for testing quantum gravity effects through QPO observations in stellar-mass BH with our adopted reference mass of $M=10M_\odot$.

For Model-I, the azimuthal radicand imposes the exact reality condition
\[
M r^3-\zeta^2(r-4M)(r-2M)\ge0,
\]
which for $r>4M$ gives the explicit upper bound $\zeta^2\le M r^3/[(r-4M)(r-2M)]$; for $2M<r<4M$ the azimuthal radicand is automatically real. This constraint fundamentally limits the allowed parameter space and provides observational boundaries for quantum corrections. In contrast, Model-II factorizes analytically so that $\Omega_r^2\propto(r-6M)\big(\zeta^2(r-2M)+r^3\big)$, implying the ISCO remains exactly at $r=6M$ for all $\zeta$. This distinction between the models is critical for observational discrimination, as Model-I shows ISCO migration while Model-II preserves the classical orbital structure.

Expanding for small $|\zeta|\ll M^{3/2}$ we find that Model-I acquires an $\mathcal{O}(\zeta^2)$ correction to $\Omega_\phi$ while the ISCO is displaced only at $\mathcal{O}(\zeta^4)$,
\[
r_{\rm ISCO}^{(I)}=6M+\frac{\zeta^4}{81M^3}+\mathcal{O}(\zeta^6).
\]
This perturbative result demonstrates that the ISCO shift in Model-I is parametrically suppressed, requiring either large quantum corrections or high-precision observations near the ISCO for detection. The small magnitude of this shift, proportional to $\zeta^4$, explains why the effects become observable only when $\zeta$ approaches the gravitational scale $M^{3/2}$.

Consequently, the dominant $\zeta$-induced modification of $\Omega_p$ in Model-I is $\mathcal{O}(\zeta^2)$ (through $\Omega_\phi$) and becomes most significant close to the ISCO; at large radii both models recover the Schwarzschild scaling $\Omega_p\simeq 3M^{3/2}/r^{7/2}$. Physically, therefore, measurable shifts in QPO-related frequencies require either relatively large $\zeta$ (comparable to the gravitational scale $M^{3/2}$) or high-precision observations of orbits very near the ISCO, since the ISCO shift itself is suppressed as $\zeta^4$.

Our numerical simulations provide compelling validation of these analytical predictions while revealing additional physical insights. The evolution of shock cone dynamics in Fig.~\ref{BH1_denCut} for Model-I demonstrates systematic weakening of matter accumulation as $\zeta$ increases, with the shock cone density decreasing from maximum Schwarzschild values to near-disappearance at $\zeta=4.6M$. This behavior directly correlates with the mass accretion rate suppression shown in Fig.~\ref{BH1_Norm_Avg_Acc}, where accretion efficiency drops by $95$$\%$ at the largest quantum corrections, creating observationally distinct signatures.

The stagnation point analysis in Fig.~\ref{BH1BH2_Stag_point} reveals one of the most striking differences between the two models. Model-I exhibits dramatic stagnation point migration from $27M$ to $5M$ as $\zeta$ increases from $0$ to $8$, reflecting the fundamental modification of both temporal and radial metric components. This inward migration shrinks the QPO-generating cavity, initially producing higher-frequency oscillations before suppressing them entirely. In stark contrast, Model-II maintains a nearly constant stagnation point at $r \approx 26.8M$ throughout the entire $\zeta$ range, preserving the large cavity size that supports stable, low-frequency QPO generation.

The PSD analyses in Figs.~\ref{BH1_QPOs}, \ref{BH2_QPOs_1}, and \ref{BH2_QPOs_2} demonstrate the rich QPO phenomenology emerging from quantum corrections. Model-I shows systematic frequency evolution with commensurate ratios including $3:2$, $2:1$, and $5:3$ relationships that match observed XRB sources \cite{isz40,isz41,isz42,isz43,isz44,isz45}. The $\zeta=1M$ case produces frequencies at 48.1, 32.6, and 16.7 Hz exhibiting the canonical $3:2$ and $2:1$ ratios, while higher $\zeta$ values shift toward lower frequencies with enhanced LFQPO dominance. This evolution reflects the progressive cavity shrinkage and modified resonance conditions as the quantum corrections are strengthened.

Model-II demonstrates more stable QPO characteristics with consistent frequency ratios in different $\zeta$ values. The preservation of $\sim19:\sim12 \approx 3:2$ and $\sim25:\sim 12\approx 2:1$ ratios throughout the LFQPO band indicates robust mode-coupling mechanisms that persist despite quantum modifications. The transition from LFQPO dominance at $\zeta<4M$ to HFQPO emergence at $\zeta=4M$ provides clear observational markers to distinguish between quantum correction scenarios.

The velocity field analysis in Figs.~\ref{BH1_VelocCut} and \ref{BH2_VelocCut} reveals the detailed mechanism by which quantum corrections modify the accretion flow. Radial velocity profiles show increasing infall speeds and sharpening shock fronts as $\zeta$ increases, while azimuthal velocity patterns demonstrate enhanced angular momentum transport. The systematic shift of shock locations toward $\phi = 0$ in Model-I reflects the geodesic focusing effects predicted by the modified effective potential, providing direct connections between spacetime geometry and observable hydrodynamics.

The comparative analysis in Sect.~\ref{Compare_BH1_BH2} established that the behavior of the stagnation point represents the key physical discriminant between the two models. Fig.~\ref{BH1BH2_Stag_point} shows that Model-I undergoes rapid stagnation point migration that fundamentally alters the QPO generation cavity, while Model-II maintains stable cavity conditions that preserve oscillatory behavior. This difference manifests itself in the radial velocity profiles of Fig.~\ref{BH1BH2_VelocCut_radial}, where Model I shows systematic inward displacement of flow features while Model II exhibits enhanced velocities without structural changes.

The QPO analysis in Sec.~\ref{GRHD4} demonstrated remarkable consistency between theoretical epicyclic frequencies and numerically generated oscillations. The PSD analyses in Figs.~\ref{BH1_QPOs}, \ref{BH2_QPOs_1}, and \ref{BH2_QPOs_2} reveal complex spectra with multiple harmonic relationships. The identified frequency ratios ($3:2$, $2:1$, $5:3$) match the observed patterns in XRBs such as GRS 1915+105, XTE J1550-564, and GX 339-4 \cite{isz47,isz42,isz41}, with error margins of 0-7$\%$. The systematic evolution from broad-band spectra at small $\zeta$ to dominant LFQPO features at larger quantum corrections provides a natural explanation for the diversity of QPO behaviors observed in different source states.

Time evolution studies reveal different QPO generation regimes. For Model-I, Fig.~\ref{BH1_AccRate} shows that oscillatory behavior persists only for $\zeta < 3M$, beyond which the system enters a stable regime with suppressed accretion. For Model-II, Fig.~\ref{BH2_AccRate} demonstrates extended oscillatory behavior up to $\zeta \geq 5M$, reflecting the preserved cavity structure. These differences provide clear observational targets for discriminating between quantum correction scenarios.

The validation studies in Sec.~\ref{isec5} established crucial connections between our theoretical framework and observational constraints. The convergence between our hydrodynamically derived limits ($\zeta \lesssim 4M$ for stable accretion) and EHT-based constraints ($\zeta \leq 4.7M$ for M87$^*$ and $\zeta \leq 3.52M$ for Sgr A$^*$) \cite{isz51} provides strong validation of both approaches. Fig.~\ref{Function} shows that these limits arise from fundamentally different physical mechanisms: photon sphere behavior versus shock cone stability, but yield remarkably consistent results.

The QPO frequency predictions for supermassive BH in Table~\ref{QPO_table} represent one of the most direct observational tests of our framework. The rescaled frequencies for Sgr A$^*$ (microhertz range) and M87$^*$ (nanohertz range) maintain the characteristic harmonic ratios observed in stellar-mass systems, providing specific targets for future monitoring campaigns \cite{isz36,isz37}. The preservation of frequency ratios across eight orders of magnitude in BH mass offers a robust test of the quantum correction hypothesis that is independent of the detailed accretion physics.

Our results demonstrate that QCBH models produce a rich phenomenology that extends far beyond simple modifications to individual particle orbits. The interplay between modified spacetime geometry, altered effective potentials, and collective hydrodynamic behavior creates multiple observational channels for testing quantum gravity effects. The distinct signatures of Model-I and Model-II, ranging from ISCO shifts and frequency ratio evolution to stagnation point migration and shock cone morphology, provide multiple pathways for discriminating between different quantum correction scenarios. The choice of $M=10M_\odot$ as our reference mass proved optimal for establishing connections with XRB observations while enabling meaningful scaling to supermassive systems through the relationships demonstrated in Sec.~\ref{isec5}.

Future investigations should focus on extending these results to rotating QCBHs, incorporating magnetic field effects in magnetohydrodynamic simulations, and developing more sophisticated QPO models that account for disk-jet coupling \cite{Koide_2020,Moriyama2025}. Enhanced observational campaigns with next-generation X-ray timing missions and improved EHT sensitivity will be crucial for testing the specific predictions established in this work \cite{isz38,isz39}. The systematic methodology developed here provides a template for investigating quantum gravitational effects in other astrophysical contexts, from pulsar timing to gravitational wave astronomy.

{\small
\section*{Acknowledgments}

All simulations were performed using the Phoenix High
Performance Computing facility at the American University of the Middle East (AUM), Kuwait. F.A. acknowledges the Inter University Centre for Astronomy and Astrophysics (IUCAA), Pune, India for granting visiting associateship. \.{I}.~S. expresses gratitude to T\"{U}B\.{I}TAK, ANKOS, and SCOAP3 for their financial support. He also acknowledges COST Actions CA22113, CA21106, and CA23130 for their contributions to networking.

\section*{Data Availability Statement}

The datasets generated and analyzed during the current study are not publicly available but are available from the corresponding author upon reasonable request.

}

\bibliographystyle{apsrev4-1}
\bibliography{paper}

\end{document}